\documentclass[
reprint,
superscriptaddress,
 amsmath,amssymb,
 aps,
 prb,
floatfix,
showkeys
]{revtex4-2}
\usepackage{xcolor}
\usepackage{graphicx}
\usepackage{dcolumn}
\usepackage{bm}
\usepackage{hyperref}

\usepackage{xr}
\usepackage{ulem}
\begin{document}


\title[]{
Modelling how lamellipodia-driven cells maintain persistent migration and interact with external barriers
}

\author{Shubhadeep Sadhukhan}
\email{shubhadeep.sadhukhan@weizmann.ac.il}
\affiliation{%
 Department of Chemical and Biological Physics, Weizmann Institute of Science, Rehovot, Israel
}

\author{Cristina Martinez-Torres}
\email{martineztorres@uni-potsdam.de}
\affiliation{Institute of Physics and Astronomy, University of Potsdam, Potsdam 14476, Germany}
\author{Samo Peni\v{c}}
\email{samo.penic@fe.uni-lj.si}
\affiliation{%
Laboratory of Physics, Faculty of Electrical Engineering, University of Ljubljana, Ljubljana, Slovenia
}
\author{Carsten Beta}
\email{beta@uni-potsdam.de}
\affiliation{Institute of Physics and Astronomy, University of Potsdam, Potsdam 14476, Germany}
\affiliation{Nano Life Science Institute (WPI-NanoLSI), Kanazawa University, Kanazawa 920-1192, Japan}

\author{Ale\v{s} Igli\v{c}}
\email{ales.iglic@fe.uni-lj.si}
\affiliation{%
Laboratory of Physics, Faculty of Electrical Engineering, University of Ljubljana, Ljubljana, Slovenia
}
\affiliation{%
Laboratory of Clinical Biophysics, Faculty of Medicine, University of Ljubljana, Ljubljana, Slovenia
}
\author{Nir Gov}%
\email{nir.gov@weizmann.ac.il}
\affiliation{%
 Department of Chemical and Biological Physics, Weizmann Institute of Science, Rehovot, Israel
}
\date{\today}

\begin{abstract}
Cell motility is fundamental to many biological processes, and cells exhibit a variety of migration patterns. Many motile cell types follow a universal law that connects their speed and persistency, a property that can originate from the intracellular transport of polarity cues due to the global actin retrograde flow. This mechanism was termed the ``Universal Coupling between cell Speed and Persistency"(UCSP). Here we implemented a simplified version of the UCSP mechanism in a coarse-grained ``minimal-cell" model, which is composed of a three-dimensional vesicle that contains curved active proteins. This model spontaneously forms a lamellipodia-like motile cell shape, which is however sensitive and can depolarize into a non-motile form due to random fluctuations or when interacting with external obstacles.  The UCSP implementation introduces long-range inhibition, which stabilizes the motile phenotype. This allows our model to describe the robust polarity observed in cells and explain a large variety of cellular dynamics, such as the relation between cell speed and aspect ratio, cell-barrier scattering, and cellular oscillations in different types of geometric confinements.
\end{abstract}

\maketitle

\textit{\textbf{Significance Statement}}---{~Coupling curved membrane proteins to active protrusive forces that arise from recruited actin polymerization, can lead, in the presence of adhesion, to self-organization of a leading-edge cluster and a motile ``minimal-cell". However, this polarized and motile shape can become unstable, and due to fluctuations or interactions with external perturbations transform to an immotile, symmetric shape. Here we couple the spatial organization of the curved active proteins to a global advection of a polarity cue along the cell's activity axis. Introducing long-range inhibition, the resultant gradient of the polarity-cue stabilizes the motile, polarized ``minimal-cell" vesicle. We thereby present a robust model of cell motility that can explain a variety of cellular shape-migration relations, cell-barrier scattering and spontaneous oscillations of confined cells.}

\section{Introduction}
During cell migration within the body, such as in development or cancer, cells often have to navigate complex geometries defined by external barriers, tissues and extra-cellular matrix (ECM) fibers \cite{yamada2019mechanisms}. These external constraints and confinements challenge the ability of cells to maintain their internal polarization and exhibit persistent migration. Cellular motility during interaction with complex external constraints and confinement presents an open challenge for our understanding of cell migration.

This process has been explored over recent years using in-vitro experiments where motile cells have been observed while migrating over various topographies and confined within various geometries \cite{ko2013directing,driscoll2014cellular,graziano2019cell,werner2019cell, Nagel2014}. Theoretical models that describe cellular migration in complex geometries have relied on different coarse-grained cell mechanics approaches \cite{kim2013dynamic,he2017substrate}, phase-field and cellular Potts model (CPM) frameworks \cite{camley2013periodic,winkler2019confinement}, or more coarse-grained approaches \cite{ron2024emergent}. These models did not consider the role of curved membrane components and local membrane curvature at the leading edge during lamellipodia-driven cell migration, which is our focus here.

Recently, we have introduced a ``minimal-cell" model where a motile phenotype emerges spontaneously due to the coupling between curved membrane complexes (CMC) and protrusive forces that result from the recruitment of actin polymerization to these membrane sites \cite{sadhu2021modelling,sadhu2023minimal}. This coupling can lead to the formation of a lamellipodia-looking protrusion, with a leading-edge cluster of the CMC, and a total resultant force that moves the vesicle in the forward direction (Fig.\ref{fig:model}A(i)). This simple model has been successful in explaining the origin of several curvotaxis behaviors \cite{sadhu2024minimal}, and since this model is based on only a few physical ingredients, it is very general and applies to many cell types. We note that there are recent experimental indications for the curvature-sensitivity of the leading-edge components of the lamellipodia \cite{linkner2014inverse,begemann2019mechanochemical,pipathsouk2021wave,wu2024wave, Baldauf2023}, as arises in our model.

Nevertheless, within this model the polarized, motile phenotype was rather fragile, and once its leading-edge cluster breaks up, the vesicle irreversibly loses its polarization and forms a non-motile two-arc shape (Fig.\ref{fig:model}B(i)). Such polarity-loss events can be triggered spontaneously by shape fluctuations or due to the vesicle interacting with external rigid barriers \cite{sadhu2021modelling}. Similar events of break-up of the leading-edge are observed in motile cells \cite{andrew2007chemotaxis,yang2011zigzag,dieterich2008anomalous}, however cells have mechanisms that allow them to repolarize and resume their motile phenotype \cite{weiner2007actin,gross2020using}.

Here we implement a mechanism for internal cellular polarization \cite{maiuri2015actin,ron2020one} into our minimal-cell model, thereby greatly increasing the robustness of the motile phenotype within this model. This allows us to use this model to explain the observed relation between cell speed and shape, as well as the scattering and spontaneous oscillations of motile cells when interacting with external rigid topographical barriers and complex adhesion patterns.

\section{Theoretical model}

\begin{figure}
    \centering
    \includegraphics[scale=0.27]{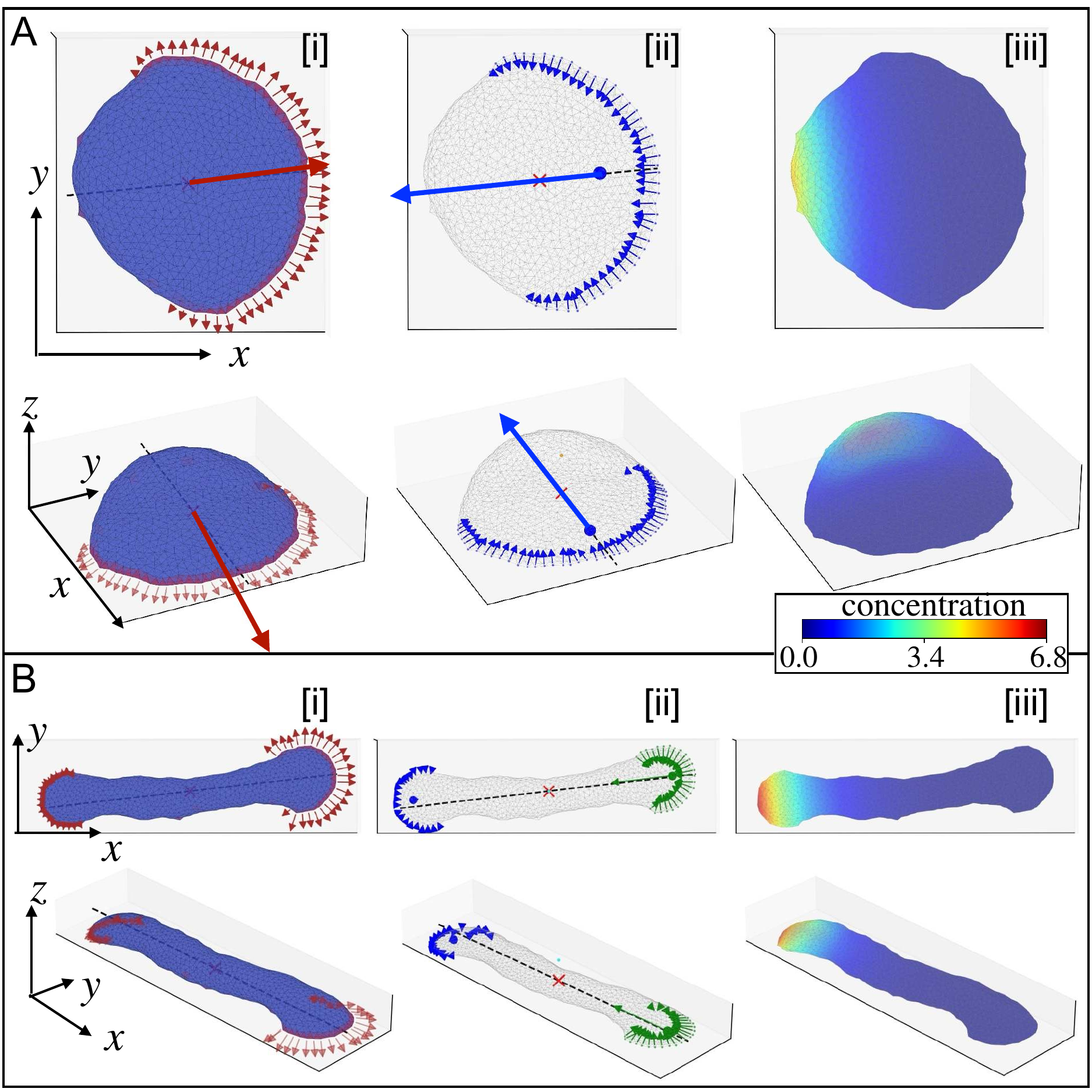}
    \caption{Implementation of the UCSP mechanism and polarity cue advection into the MC simulations of the vesicle shape. This is demonstrated here for two main phenotypes that coexist in the absence of UCSP \cite{sadhu2021modelling}: (A) Polar crescent-shaped vesicle ($E_{\rm ad}=3k_BT,~F=2 k_BT l_{\rm min}^{-1}$): [i] The active force at each CMC site is denoted by a red arrow, directed at the local outwards normal. The big red arrow at the centre of mass (denoted by a red cross) denotes the total active force due to all the active CMC. The black dashed line shows the calculated axis for the net internal flow. [ii] The centre of mass of the protein cluster is denoted by a blue solid circle. The small blue arrows denote the local contribution of each CMC to the global actin retrograde flow (opposite to the direction of the local active force in [i]), and the large arrow denotes the total internal actin-retrograde flow (Eq.\ref{eq:vmag}). [iii] The concentration profile of the polarity-cue (inhibitor of the actin polymerization activity), using Eq.(\ref{eq:threeD_formula}).
    (B) Non-polar and immotile two-arc-shaped vesicle ($E_{\rm ad}=1k_BT,~F=3 k_BT l_{\rm min}^{-1}$).[i]-[iii] as in (A). For each snapshot, we present top and side views.}\label{fig:model}
\end{figure}

Our minimal-cell model is based on the Monte-Carlo calculation of the dynamics of a closed three-dimensional triangulated self-avoiding vesicle with a spherical topology (Fig.\ref{fig:model})~\cite{Fonari2019,sadhu2021modelling,sadhukhan2023modelling} (See SI section A for details). Within this model we denote the bare membrane nodes in blue, and the nodes containing the CMC in red (where the active protrusive forces are applied).

The mechanism we implement for the internal polarization of the cell is based on the Universal Coupling between cell Shape and Persistency (UCSP)~\cite{maiuri2015actin,ron2020one,ron2024emergent}. Within this model the actin polymerization at the leading edges of the cell gives rise to a net advection of polarity cues across the cell. This advection results in a gradient of the polarity cue along the cell length, and these polarity cue gradients in turn affect the actin polymerization activity at the leading edges, thereby completing a positive feedback that can give rise to spontaneous polarization of the cell. The model was previously mostly implemented in a simple one-dimensional representation of the cell ~\cite{maiuri2015actin,ron2020one,ron2024emergent}, and here we similarly implement it in a simplified manner by projecting the polarity cue gradient and feedback along a one-dimensional axis (ignoring hydrodynamic flow patterns - see in a two-dimensional cell \cite{lavi2020motility}). We further note that the front-back inhibition may arise in cells due to other types of long-range inhibition, and one can view our implementation as an example of a large class of intra-cellular mechanisms that stabilize cell polarization \cite{weiner2002regulation,mori2008wave,houk2012membrane,goehring2013cell, Rappel2017, Illukkumbura2020}. 

First, we calculate the axis of the net retrograde flow within the vesicle, for the instantaneous configuration of membrane shape and CMC clusters.
This is shown in Fig.\ref{fig:model}, and more details are given in SI section B.
The magnitude of the net retrograde flow is related to the active cytoskeleton force exerted by the CMC (Fig.\ref{fig:model}(A,B)[i]) by a positive coupling factor $\beta$. Each CMC contributes to the total actin retrograde flow, which is oppositely directed to the active protrusive force
\begin{equation}
    V=-\beta\sum_i \boldsymbol{F}^{\rm act}_i\cdot\hat{\boldsymbol{e}},
    \label{eq:vmag}
\end{equation}
where, $\hat{\boldsymbol{e}}$ is the direction of net retrograde flow along the calculated axis in such a way as to have a positive $V$ (Fig.\ref{fig:model}(A, B)[ii]), and $\boldsymbol{F}^{\rm act}_i$ is the active actin-driven protrusive force due to a CMC at the $i$th vertex, pointing at the outwards normal.

Using the calculated retrograde flow, we calculate the polarity cue distribution along this axis, given by,
\begin{equation}
    c(p)=\frac{c_{\rm tot}V (p_f-p_b)}{D V_{\rm ves}}\frac{e^{-Vp/D}}{e^{-Vp_b/D}-e^{-Vp_f/D}},
    \label{eq:threeD_formula}
\end{equation}
as shown in (Fig.\ref{fig:model}(A,B)[iii]). Here, $p$ is the projection of a vertex position with respect to the centre of mass on the axis of retrograde flow, $V_{\rm ves}$ is the volume of the vesicle, and $D$ is the diffusion coefficient of the inhibitory polarity cues. $p_f$ and $p_b$ are the projections of the vertices at the front and the back of the cell with respect to the retrograde flow axis. The quantity $c_{tot}$ denotes the total amount of the polarity cue:
$c_{\rm tot}=\int_{\rm vesicle} c(p) d\tau$. In our model the unit of the actin retrograde flow $V$ is given by $D/l_{\rm min}$, since its only the ratio $V/D$ that appears in Eq.\ref{eq:threeD_formula}. The force has units of $k_BTl^{-1}_{\rm min}$, and therefore the coupling parameter $\beta$ has units of $\frac{D}{l_{\rm min}}\frac{1}{k_BT l^{-1}_{\rm min}}=D/k_BT$.

The polarity cue distribution in turn affects the local strength of the protrusive force induced by actin polymerization at the location of each CMC, according to the following steady-state of first-order kinetics
\begin{equation}
    \tilde{F}=\frac{F}{1+c(p)/c_s},
    \label{eq:reduced_force}
\end{equation}
where $c_s$ is the saturation concentration for this inhibitory reaction. This feedback means that in the region with the higher concentration of inhibitor the CMC induce weaker protrusive forces, and contribute less to the total actin retrograde flow (Eq.\ref{eq:vmag}). This is clearly demonstrated in Fig.\ref{fig:model}B.

The modified actin polymerization forces (Eq.\ref{eq:reduced_force}), and the modified contribution to the actin retrograde flow (Eq.\ref{eq:vmag}), are then used to calculate again the polarity cue profile (Eq.\ref{eq:threeD_formula}) until these repeated iterations converge to within some fixed error threshold of $0.1\%$. The MC simulation is then allowed to proceed with the modified CMC forces, until the vesicle shape has changed by more than some threshold value $10\%$, and the UCSP calculation is updated, whereby a new axis for the net flow is calculated and the whole process is repeated.

The examples shown in Fig.\ref{fig:model} demonstrate the principles of this procedure for two typical shapes that form spontaneously in our model: the motile crescent shape and the non-motile two-arc shape. Both shapes co-exist in the same parameter regime, when we do not implement the UCSP calculation. In both cases, we indicate the identification of the axis of the net retrograde flow, and the resulting concentration profile of the polarity cue. We show here the first iteration of the UCSP calculation.

In the UCSP model we found that spontaneous polarization of the cell occurs when the coupling parameter between the actin flow and the asymmetry in the polarity cue ($\beta$) is larger than a critical value \cite{maiuri2015actin}, which also depends on the cell length \cite{ron2020one}. We show in the SI a similar transition from a non-motile to a motile (polarized) state above a critical value of $\beta$ (Fig.~S-1).

All the vesicles used in this work are composed of $N=1447$ vertices to create the triangulated membrane. We used the time unit as 20000 Monte Carlo steps and the protein percentage is $\rho=3.45\%$ if not specified (For pancake shape, we used $\rho=5.53\%$ in Fig.\ref{fig:traj_blow}). We used the following UCSP parameters throughout the paper: $c_{\rm tot}=4000$, $D=4000$, and $c_s=1$, and the CMC binding energy was fixed at $w=1~k_BT$.

\section{Results}

We now demonstrate the results of implementing the internal polarization mechanism (UCSP) on the motility dynamics of our minimal-cell model.
\begin{figure*}
    \centering
    \includegraphics[width=1\linewidth]{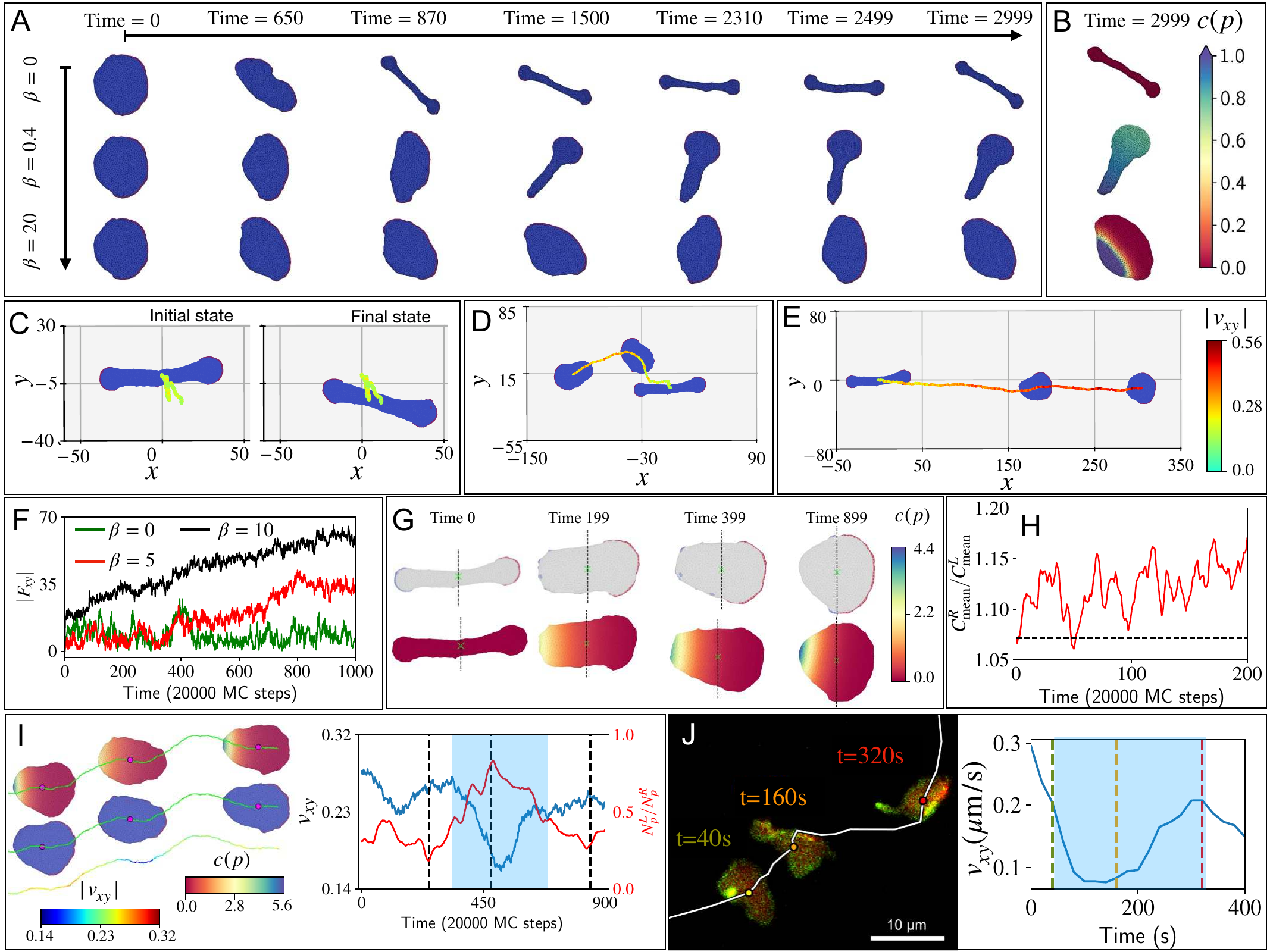}
    \caption{Stabilization of the crescent-shaped, polar vesicle by the UCSP mechanism. (A) Examples of the dynamics of the motile vesicle for different values of the UCSP coupling strength $\beta$. In the absence of UCSP ($\beta=0 ~D/k_BT$) the crescent shape is transiently stable, and thermal fluctuations break it into the two-arc shape \cite{sadhu2021modelling}. As the coupling strength $\beta$ increases, the instability is delayed or allows the polar state to be partially recovered ($\beta=0.4~D/k_BT$). When the coupling is strong ($\beta=20~D/k_BT$) the crescent shape becomes absolutely stable. In order to maintain a constant maximal active force magnitude of $\tilde{F}_{max}\approx 4k_BT l_{\rm min}^{-1}$,  we used $F=4,~7.1,~4k_BT l_{\rm min}^{-1}$ for the cases of $\beta=0,~0.4,~20.0 ~D/k_BT$ respectively (the adhesion parameter $E_{\rm ad}=1~k_BT$). See Movie~S-1. (B) The concentration profile of the polarity cue along the UCSP axis at the final times of the cases shown in (A). (C-E) The transition of a two-arc shape to a crescent polar-shaped vesicle (See Movie~S-4). The trajectories indicate the speed of the vesicle using the heatmap. We started the simulations with a two-arc-shaped vesicle using $F=3 k_BT l_{\rm min}^{-1}$, $E_{ad}=1~k_BT$. (C) In the absence of UCSP coupling ($\beta=0~D/k_BT$), the two-arc shape is stable. (D,E) For higher coupling strength ($\beta=5,~10~D/k_BT$), the UCSP mechanism breaks the symmetry of the two-arc shape and makes a transition to the crescent, motile shape (shown at times $0, ~699,~999$). (F) The planar force magnitude $F_{xy}$ is shown for $\beta=0,~5,$ and $10$ in units of $D/k_BT$ with green, red and black respectively. (G) The snapshots of the shape transition from a two-arc to a crescent shape for $\beta=10~D/k_BT$ (the case $\beta=10~D/k_BT$ shown in E). The CMC on the right ($x_i>x_{\rm CM}$) and left ($x_i<x_{\rm CM}$) of the centre of mass are marked by red and blue respectively. The concentration profile of the inhibitory polarity cues is shown. (H) The time evolution of the ratio of the average curvature of the CMC on the right and left of the centre of mass. (I) Increased amplitude of thermal fluctuations by decreasing the bending rigidity $\kappa=15 k_BT$, giving rise to spontaneous transition between polar and non-polar shapes. We set the coupling parameter $\beta=20~D/k_BT$. The polar-shaped vesicle makes a transition to a nearly two-arc shape and back to a motile crescent shape (See Movie~S-3). The two-arc shape corresponds to the dip in the planar speed plot (blue), and a peak in the ratio of left/right number of CMC with respect to the c.o.m. along the polarity axis (red). (J) Experimental data is showing a \textit{D. discoideum} cell undergoing a transition between polar and non-polar (cells scaled down by a factor of $2$, labelled with LifeAct-GFP and PHcrac-RFP). The snapshots are shown at 40s, 160s and 320s (See Movie~S-5). The planar speed $v_{xy}$ shows a similar dip when it becomes a two-arc. The dashed lines correspond to the time at which we showed the shape of the cell from the experiment.}\label{fig:shape_transition}
\end{figure*}
 
\subsection{UCSP mechanism stabilizing the motile phenotype}

We have previously shown that the crescent shape is transiently stable and can make a spontaneous transition to a two-arc shape, since both shapes coexist in the same parameter regime \cite{sadhu2021modelling}. This instability occurs faster for large protrusive force $F$ and weak adhesion $E_{\rm ad}$, and is shown in Fig.\ref{fig:shape_transition}. In Fig.\ref{fig:shape_transition}A we demonstrate that in the absence of UCSP ($\beta=0~D/k_BT$), the leading-edge cluster can spontaneously break into two, whereby the vesicle changes to the two-arc shape and motility is irreversibly lost. As the coupling between the asymmetry in the polarity cue and the actin flow ($\beta$) increases, the UCSP mechanism can stabilize the crescent-shaped vesicle and suppress the transition to the non-motile two-arc shape. As $\beta$ increases the polarity cue profile becomes sharper from back to front (Fig.\ref{fig:shape_transition}B), and the net retrograde flow remains more stable (Fig.~S-2), preventing the transition to the two-arc shape.





Next, we study the UCSP-induced polarization of the two-arc shape (Fig.~\ref{fig:shape_transition}). We start the simulation with a non-motile two-arc-shaped vesicle, which is stable in the absence of the UCSP mechanism (Fig.~\ref{fig:shape_transition}C). We then switch on the UCSP mechanism and demonstrate how the vesicle becomes more polarized, crescent-like and motile as $\beta$ increases (Fig.~\ref{fig:shape_transition}D-E). The total active force increases as the cell polarizes and transforms from the two-arc to the crescent shape, as shown in Fig.~\ref{fig:shape_transition}F.

The polarization process proceeds as follows: the polarity cue concentration peaks at one end of the two-arc shape (Fig.~\ref{fig:shape_transition}G), inhibiting the protrusive forces in this ``losing" CMC cluster. This reduction in the amplitude of the protrusive forces causes this cluster to lose its high curvature (Fig.~\ref{fig:shape_transition}H), which then loses its stability, breaks up and its CMC diffuse to join the "winning" cluster at the opposite end of the two-arc shape (Fig.~\ref{fig:shape_transition}G), where the large protrusive force maintains a high curvature at the leading edge. This process thereby converts the two-arc shape into the crescent motile phenotype with a single large CMC cluster (Fig.~\ref{fig:shape_transition}D,G). Clearly, there is a transition regime of values $\beta\sim 1-5$ in units of $D/k_BT$ above which the minimal-cell can robustly repolarize following its transition to the non-motile two-arc shape. 

In order to observe more dynamic transitions in polarity, which are often observed in living cells, we need to allow for larger fluctuations in our system. By reducing the bending rigidity parameter from $\kappa=20~k_BT$ to $\kappa=15~k_BT$ (and using an active force parameter $F=2k_BTl_{\rm min}^{-1}$) we induce larger amplitude shape fluctuations, effectively emulating the large level of metabolic noise observed in cells. Now, we observe spontaneous shape changes (See Movie~S-3) from crescent-shaped polar to non-polar nearly two-arc-shaped vesicle, as shown in Fig.~\ref{fig:shape_transition}I. The planar speed $v_{xy}$ dips when the vesicle transitions to a two-arc shape (blue line in Fig.~\ref{fig:shape_transition}I). To quantify the distribution of the CMC clusters we plot the ratio $N^L_p/N^R_p$ of the number of proteins on the left and right sides of the centre of mass, along the polarity axis  (red line in Fig.~\ref{fig:shape_transition}I). As this ratio approaches $0$ ($1$) it implies a more (less) polar vesicle, and indeed this ratio $N^L_p/N^R_p$ reaches its maximum when the planar speed is lowest. This sequence of polarity loss, and recovery, is observed in experiments with \textit{D. discoideum} cells, where spontaneous switches between a crescent-shaped polar mode of locomotion and a non-polar state were reported~\cite{Moldenhawer2022} for an example see  Fig.~\ref{fig:shape_transition}J (See Movie~S-5). A similar dip in planar speed $v_{xy}$ is observed experimentally when the cell takes a two-arc shape.  Note that in the absence of UCSP ($\beta=0~D/k_BT$), the larger thermal fluctuations break up the motile vesicle, which quickly becomes two-arc (see Movie~S-2).

\begin{figure}
    \centering
    \includegraphics[scale=0.24]{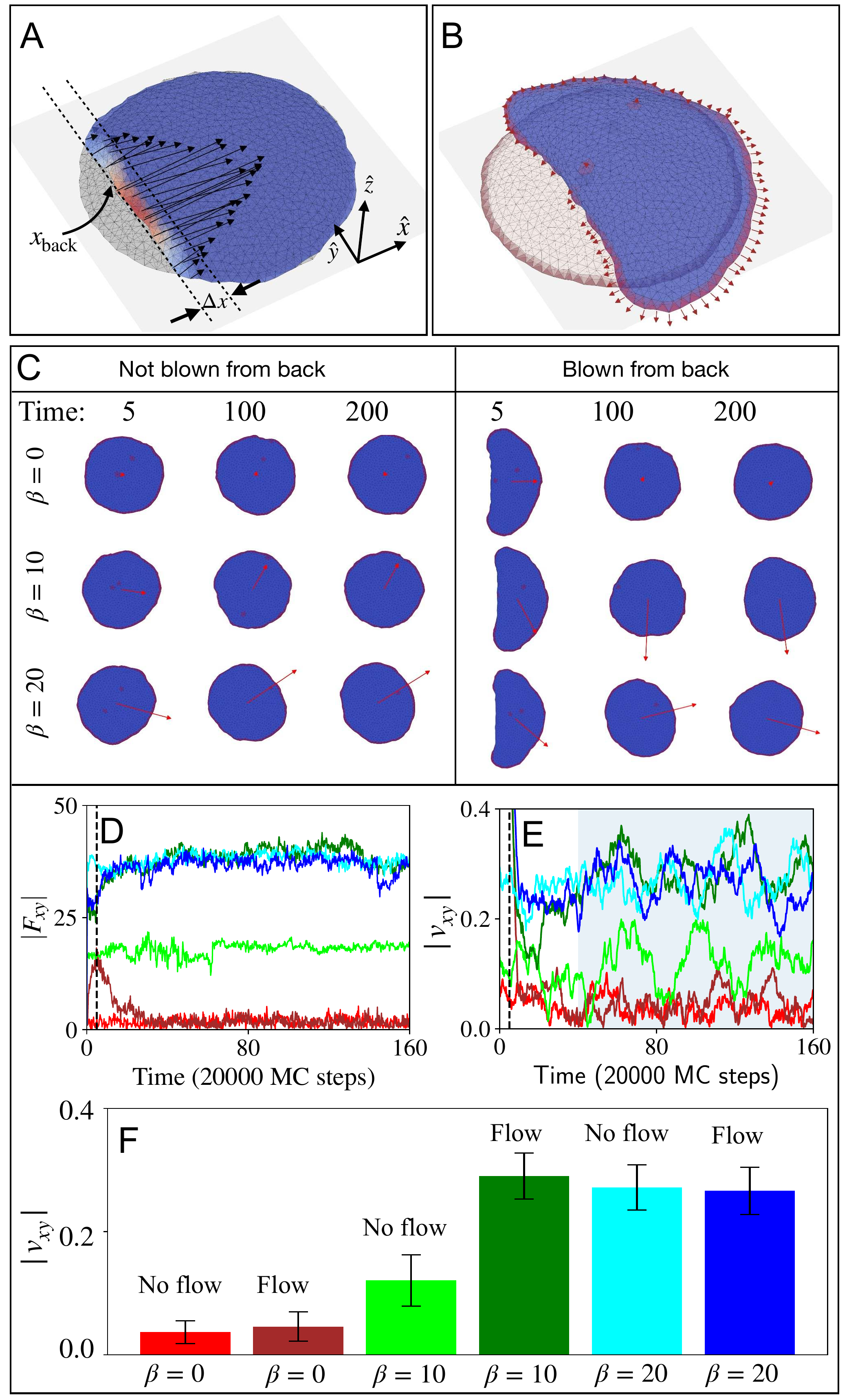} 
    \caption{The relation between the vesicle's initial shape and polarization. 
    (A,B) The initial state of a spread vesicle, with a circular leading-edge cluster (grey background image). An external force field (black arrows in (A)) is applied to deform the vesicle, mimicking the effect of a strong fluid flow in the $\hat{x}$ direction. The force is acting within a range of $\Delta x$ of the vesicle's rear. The strength of the force has a Gaussian form (heatmap). (B) The final deformed, crescent shape when the external force field is turned off. The leading-edge cluster is broken-up in the region deformed by the external force, due to the membrane losing its high curvature. The local active forces are indicated by the red arrows. (C) Vesicle shapes at different times for different coupling strengths ($\beta=0,~10,20$ in units of $D/k_BT$), for the pancake and the deformed crescent shapes (B). (D,E) The time dependence of the total planar active force $F^{\rm tot}_{xy}$ and velocity $v_{xy}$. The colour code of the lines is elaborated in (F), where we showed a bar plot of $v_{xy}$ averaged over a time window indicated by the shaded region in (E). The parameters used: parameter protein density $\rho=5.53\%$, adhesion strength $E_{\rm ad}=3k_BT$ and active force parameter $F=2,~2.5,$ and $2$ in units of $k_BT l^{-1}_{\rm min}$ for the cases $\beta=0,~10,$ and $20$ in units of $D/k_BT$ respectively to maintain the maximum force at one vertex $\tilde{F}_{\rm max}\approx 2k_BT l^{-1}_{\rm min}$. }\label{fig:traj_blow}
\end{figure}

We further investigated the interplay between cell shape and polarization by starting with a non-polar vesicle where the CMC form a circular leading edge around the entire shape (using a higher CMC concentration, Fig.\ref{fig:traj_blow}A). In the absence of UCSP this ``pancake" shape is stable and non-motile \cite{sadhu2021modelling}. We then apply a transient external force field, which emulates the effect of blowing fluid at high pressure on the vesicle (see SI section E for the details of this external force field), resulting in the deformation of the pancake into a crescent shape (Fig.\ref{fig:traj_blow}B). This simulation is motivated by the experiments that demonstrated the conversion of a non-motile circular cell fragment into a motile phenotype by deforming it into a crescent shape by an applied shear flow \cite{verkhovsky1999self}.

In Fig.\ref{fig:traj_blow}(C-F) we demonstrate the effect of such a transient shape deformation on the polarization of the vesicle, for different values of the UCSP coupling parameter $\beta$. When the transient external force field is switched off, we implement the UCSP mechanism, and follow the consequent evolution of the vesicle's polarization. We find that the shape deformation leads to a small polarization for small (or no) UCSP coupling strength, which decays over time. Initially the crescent shape disrupts the CMC cluster in the region facing the external force field (Fig.\ref{fig:traj_blow}B), as its curvature becomes small or negative (concave). However, this polarization of the CMC and the total force is short-lived and the vesicle resumes the pancake shape, with decaying polarization. 

At large values of $\beta$ the UCSP coupling is strong enough to break the symmetry and polarize even the pancake shape without the transient crescent shape deformation. There is however an intermediate regime (for example $\beta=10~D/k_BT$), where we find that the transient crescent shape deformation enables the system to attain a persistent high polarity form, which it can not reach spontaneously from the pancake shape. This observation demonstrates that within our model there can be coexistence of long-lived low and high polarity phenotypes which depend on their initial shape, as observed in experiments \cite{verkhovsky1999self}.

\begin{figure}
    \centering
    \includegraphics[scale=0.53]{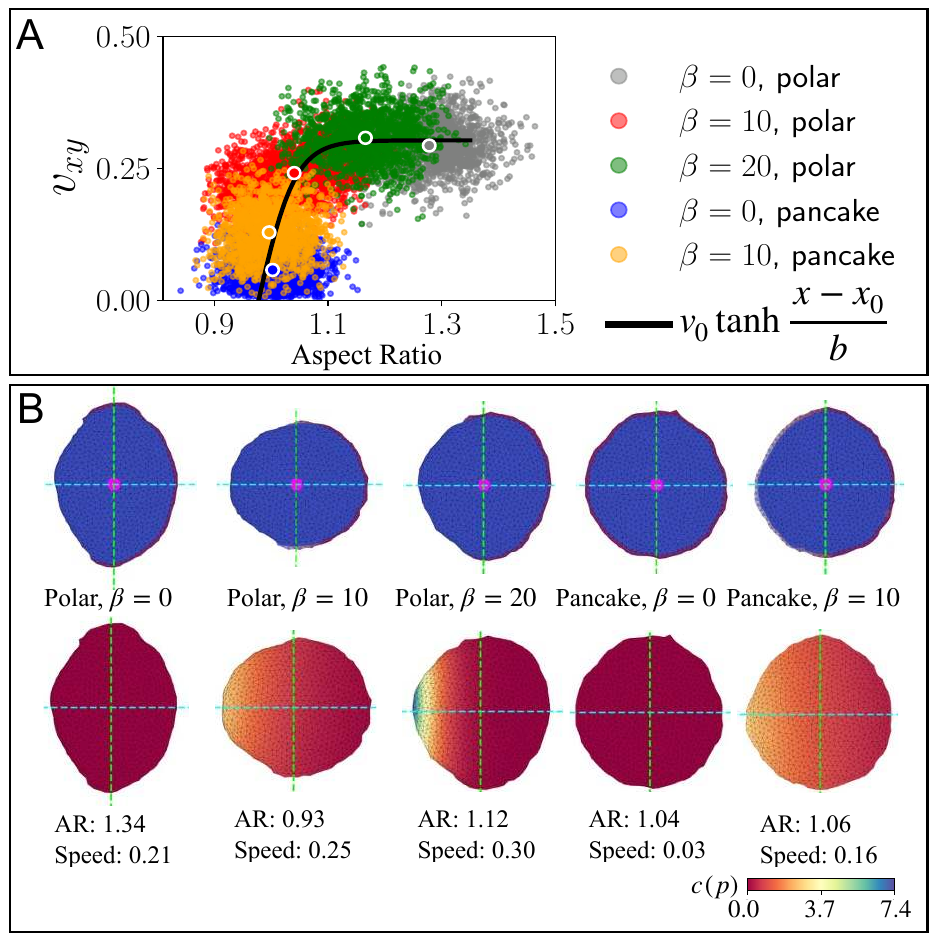} 
    \caption{Relation between the vesicle's aspect ratio and speed. (A) Scatter plot of the instantaneous speed and aspect ratio of the vesicle for four different cases: (i) polar vesicle, $\beta=0~D/k_BT$, (ii) polar vesicle, $\beta=10~D/k_BT$, (iii) polar vesicle, $\beta=20~D/k_BT$, (iv) pancake-shaped vesicle, $\beta=0~D/k_BT$ in grey, red, green, and blue respectively. The black line shows the fitted-curve with a saturating functional form $v_0\tanh{b(x-x_0)}$ through the average coordinates of each cluster. We find $v_0=0.3\pm 0.0014,~b=0.06\pm 0.002,$ and $x_0=0.977\pm 0.0003$ using python scipy package. (B) Typical shapes of the vesicles for the different cases, and the polarity cue distribution profiles. The CMC density is 3.45\% and 5.53\% for the polar and the pancake shapes respectively.}\label{fig:aspect-ratio}
\end{figure}

The correlation between cell polarization, as manifested by the cell velocity, and cell shape were experimentally measured for different types of motile cells \cite{keren2008mechanism,raynaud2016minimal}. We use our model to explore this relation by simulating vesicles with different densities of CMC and strength of the UCSP coupling (Fig.\ref{fig:aspect-ratio}). For each combination of CMC concentration and $\beta$ we plot in Fig.\ref{fig:aspect-ratio}A the instantaneous speed and aspect ratio of the vesicle's projection on the $x-y$ plane (calculated with respect to the direction of motion).  In Fig.\ref{fig:aspect-ratio}(B-E) we show typical snapshots of the vesicle shape and CMC cluster along the vesicle leading edge. Remarkably, the relation that we obtain between speed and aspect ratio (Fig.\ref{fig:aspect-ratio}A) exhibits the same step-like behaviour observed in experiments, including the large scatter \cite{keren2008mechanism,raynaud2016minimal}.

\begin{figure}
    \centering
    \includegraphics[width=1\linewidth]{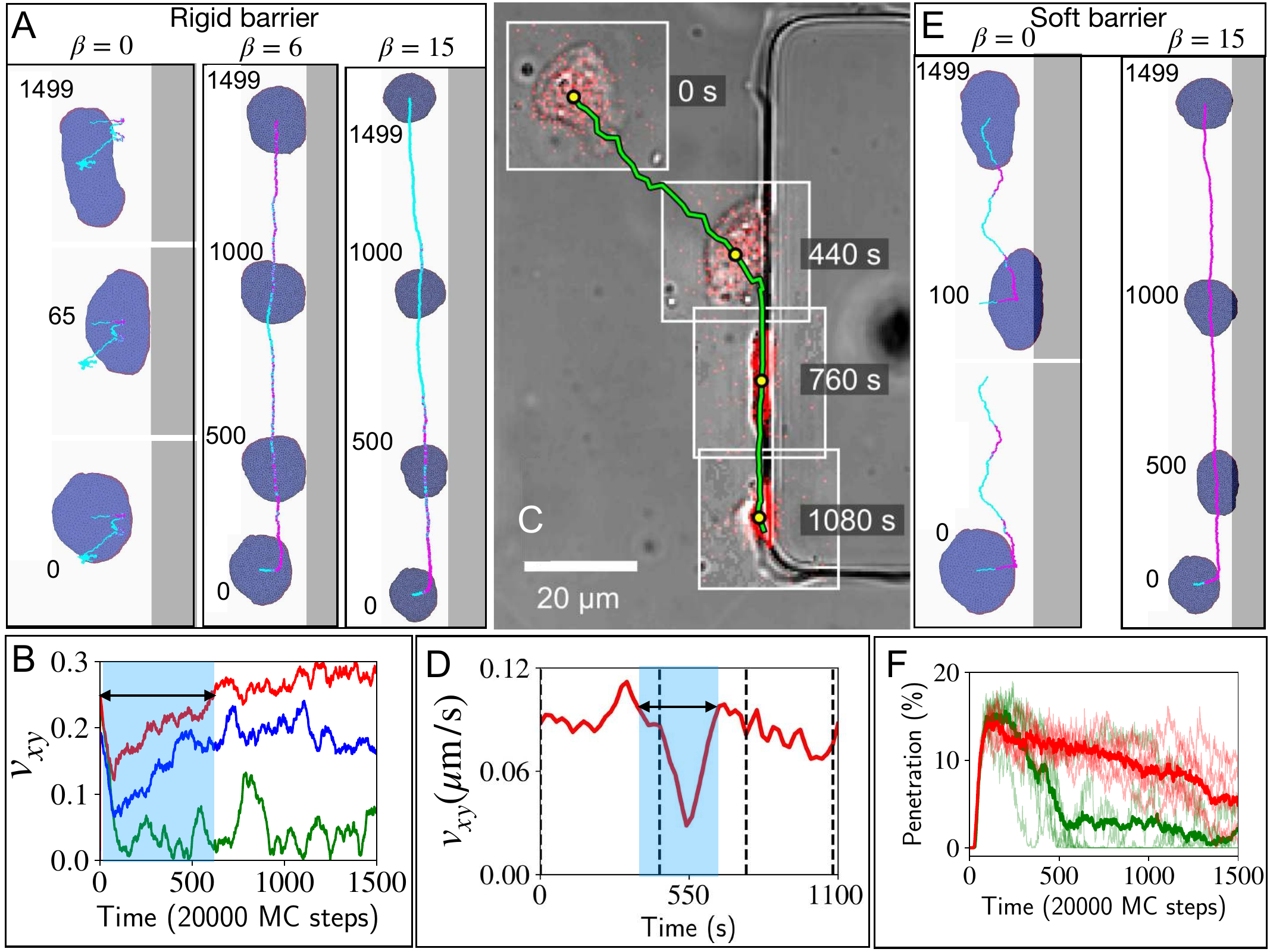} 
    \caption{Scattering and repolarization of a cell hitting a rigid barrier. (A) The trajectory and shapes of the vesicle hitting a rigid barrier (grey region) for three different values of the coupling parameter $\beta=0,~6,~15$ in units of~$D/k_BT$ (See Movies~S-6,~S-7). (B) The planar speed $v_{xy}$ from the simulation is shown for the rigid barrier. We used green, blue and red colours for $\beta=0,~6,~15$ in units of $D/k_BT$ respectively. (C) Trajectory and shapes of a \textit{D. discoideum} cell migrating and hitting a rigid barrier in the experiment (see Movie~S-9). The snapshots from different times (insets with a white border) where overlaid on the bright field channel, to visualize the barrier. Only the red channel (PHcrac) and bright field channel are shown. (D) The planar speed $v_{xy}$ of the \textit{D. discoideum} cell from the experiment. (E) The trajectory and shapes of the vesicle hitting a soft barrier (spring constant $k_w=0.1 k_BT l_{min}^{-2}$, grey region) for two values of the coupling parameter $\beta=0,~15$ in units of $D/k_BT$. (F) Penetration percentage is plotted over MC steps in the case of soft barriers for $\beta=0,~15~D/k_BT$ in green and red respectively. The trajectories in (A) and (E) are coloured cyan if the vesicle is not in contact with the wall, or magenta if it is touching (within $s=0.15~l_{min}$ of the rigid wall) or penetrating the soft wall region.  We used $F=2k_BT~l_{\rm min}^{-1}$, $E_{ad}=3k_BT$.}\label{fig:soft_vs_hard_wall}
\end{figure}

As expected, the vesicles with the lower concentration of CMC maintain a more polar form, with and without the UCSP. In fact, the vesicles with UCSP of intermediate coupling strength ($\beta=10~D/k_BT$, red dots) have a lower aspect ratio compared to no UCSP ($\beta=0~D/k_BT$, grey dots) since the forces exerted sideways, which stretch the vesicle perpendicular to its direction of motion, are inhibited in this case. With larger aspect-ratio a larger proportion of the forces are oriented in the direction of motion, allowing for higher average speed. When the CMC form a circular cluster, the cell can still polarize for significant coupling strength ($\beta=10~D/k_BT$, similar to Fig.\ref{fig:traj_blow}), but with lower speeds due to significant forces acting opposite to the direction of motion. 

Our model therefore allows to explain the aspect-ratio-speed relation, through the processes by which the active forces at the leading edge both propel and deform the cell, with the efficiency of the propulsion (speed) dependent on the shape and the resulting distribution of the leading-edge cluster around the cell. Note that our model at present does not include the contractile forces that occur at the rear of polarized cells and cell fragments \cite{verkhovsky1999self,keren2008mechanism,raynaud2016minimal, Ghabache2021}. Such contractile forces, localized at the cell rear, can deform polarized cells and cell fragments into the crescent shapes observed in experiments.

\subsection{Interactions with barriers and confinements}

We now explore the dynamics of our minimal-cell model, incorporating the UCSP, when interacting with external barriers and confinements. In the absence of the UCSP ($\beta=0~D/k_BT$), we found that when our motile vesicle impinges on a rigid wall barrier, it loses its polarity and converted to the non-polar two-arc shape \cite{sadhu2021modelling} (Fig.\ref{fig:soft_vs_hard_wall}A). This happens in our model due to the membrane flattening against the rigid barrier, losing its high curvature along the leading-edge and the subsequent migration of the highly curved CMC to the nearest free membrane on either side along the barrier (See Movie~S-6). While real cells do transiently lose or diminish their polarity when scattering off barriers, they can recover their motility and migrate away \cite{weiner2007actin,gross2020using}. Similar behaviour is observed when cells collide \cite{sitarska2023sensing}, giving rise to cell-cell scattering and a form of contact inhibition of locomotion (CIL) \cite{roycroft2016molecular}.

\begin{figure}
\centering
\includegraphics[scale=0.2]{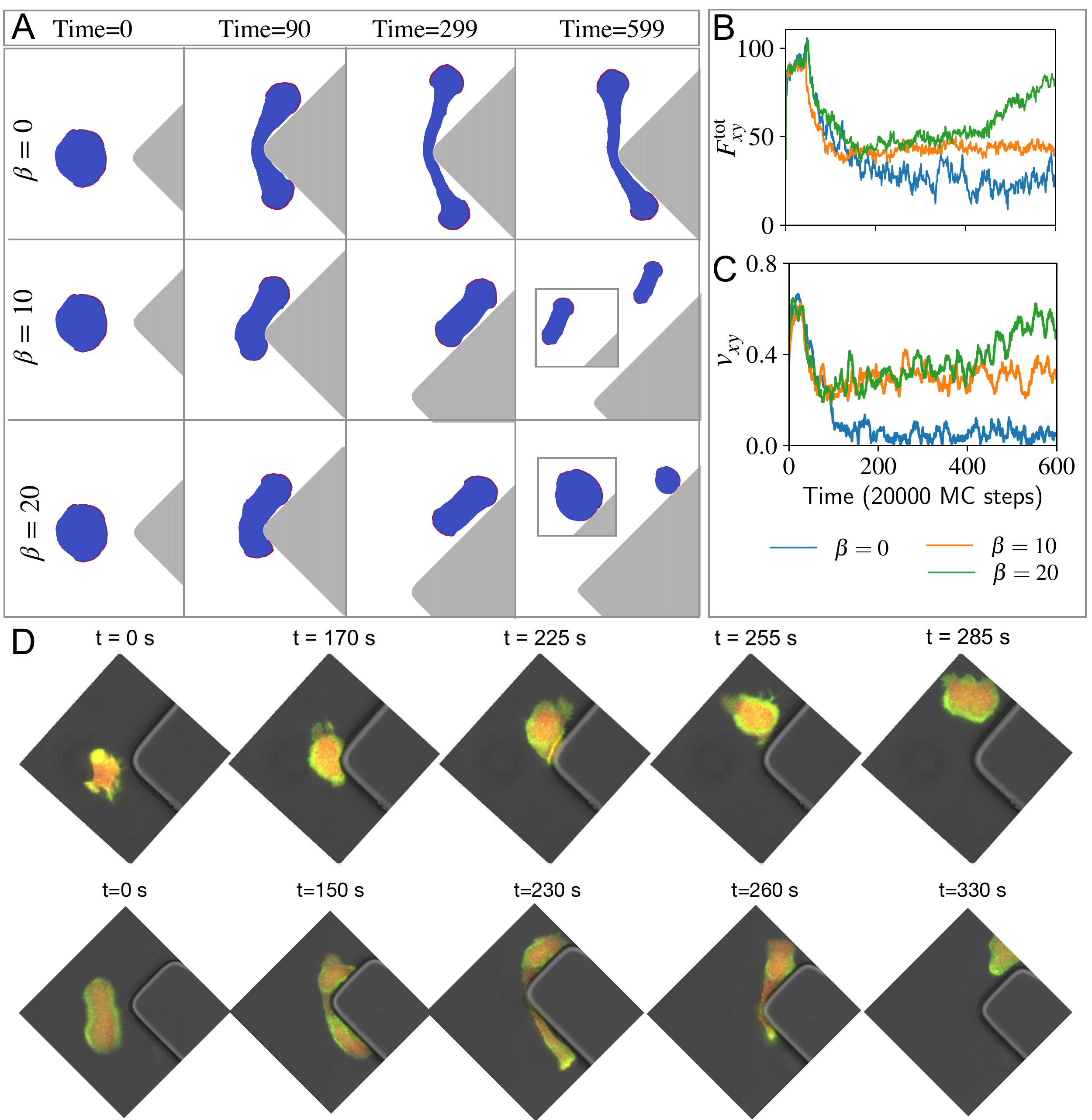}
\caption{Scattering of a vesicle and a cell from a triangular shape. (A) Snapshots of the vesicle, initially in the polar state, when hitting head-on a rigid boundary with a triangular shape, for four different coupling strengths $\beta=0,~10,~20~D/k_BT$. We used the active force ${F}=4 k_BT l_{\rm min}^{-1}$ and adhesion strength $E_{\rm ad}=3 k_BT$. For $\beta=0~D/k_BT$, we can see that the vesicle loses its polarity. For intermediate coupling, $\beta=10~D/k_BT$, the repolarization of the vesicle is partial, while it is complete for strong coupling, $\beta=20~D/k_BT$. (B,C) The total planar active force $F^{\rm tot}_{xy}$ and velocity $v_{xy}$, respectively. This demonstrate the loss of polarization, and its recovery as function of time for the different values of $\beta$. (D) Snapshots from experiments showing similar dynamics of motile \textit{D. discoideum} cells scattering off the triangular tip of a PDMS barrier. The cropped regions of interest (ROIs) have been rotated to display the same orientation as the examples shown in (A). The ROI in the first and second row is 50$\times$ 50 ${\rm \mu m}$ and 75$\times$ 75 ${\rm \mu m}$ respectively.}\label{fig:tip}
\end{figure}

In Fig.\ref{fig:soft_vs_hard_wall}A we demonstrate the trajectory of the same vesicle in the presence of UCSP ($\beta=6,15~D/k_BT$) (See Movie~S-7 for $\beta=15~D/k_BT$). While the speed and polarization transiently diminish when the vesicle hits the barrier (Fig.\ref{fig:soft_vs_hard_wall}B, Fig.~S-5A), it recovers and the vesicle migrates away. In Fig.~S-6 we give more examples of cell-barrier scattering simulations at various angles.  

In Fig.\ref{fig:soft_vs_hard_wall}C, we demonstrate an experimental trajectory of a \textit{D. discoideum} cell that hits a rigid barrier, and continues to slide along it. The speed of the cell is shown in Fig.\ref{fig:soft_vs_hard_wall}D. During the interaction with the barrier, the cell transiently loses its motility (blue shaded region) and then recovers the speed again as it slides along the barrier edge. This is very similar to the behavior in our simulations (Fig.\ref{fig:soft_vs_hard_wall}A,B).

Recently it was observed that when the cells can partially penetrate the barrier, they often get trapped at the barrier for a significant period of time, before escaping away \cite{truszkowski2023primordial}. We simulate the effect of a soft barrier by allowing the vesicle to move into the barrier (Fig.\ref{fig:soft_vs_hard_wall}E, F), which exerts a spring-like restoring force on each membrane node, proportional to the penetration distance (see SI section F for more details). 

\begin{figure}
    \centering
    \includegraphics[scale=0.24]{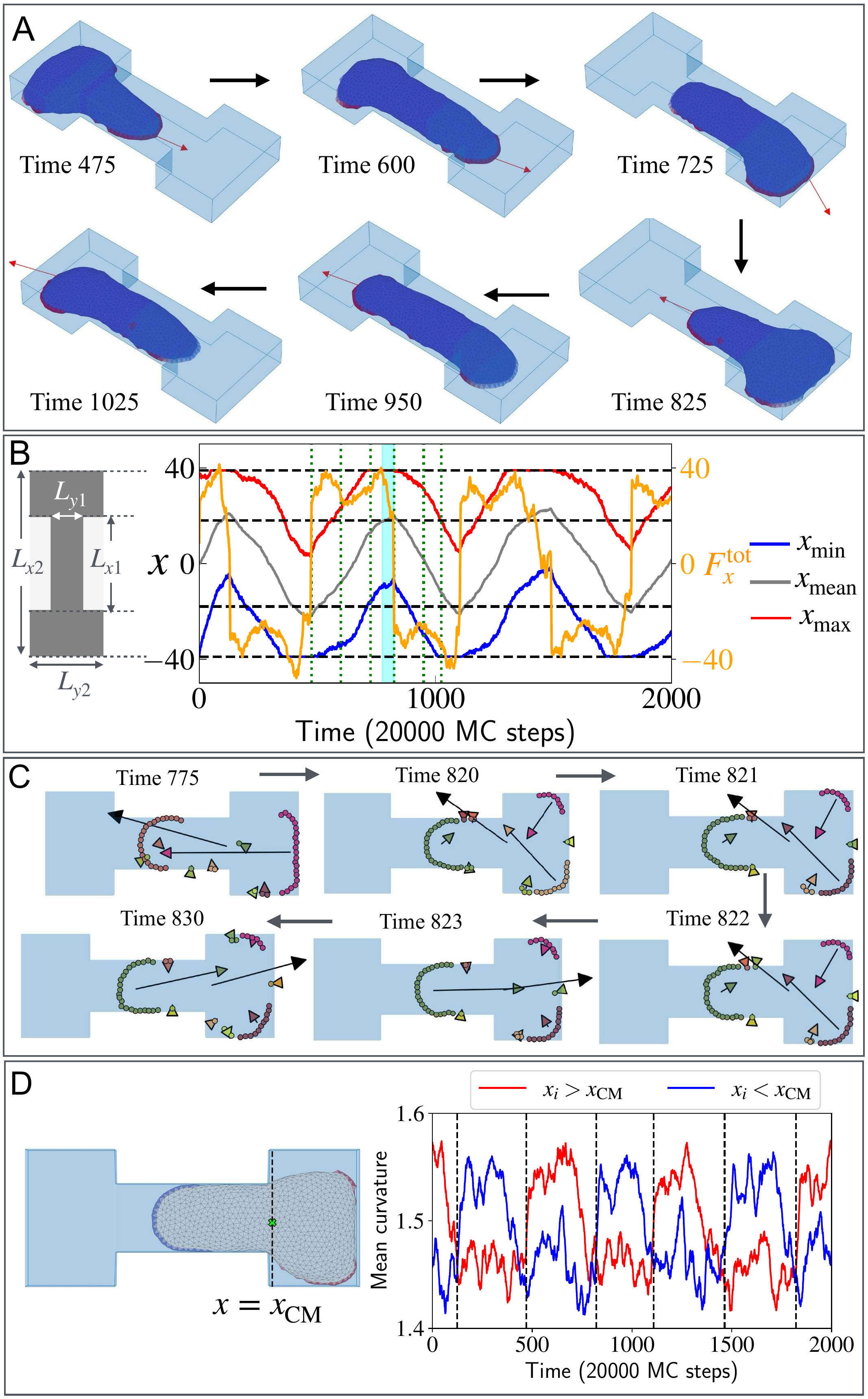}
    \caption{Oscillations within a dumbbell topographic confinement. (A) Snapshots of the vesicle within the dumbbell-shaped confinement at time 475, 600, 725, 825, 950, and 1025 in the units of 20000 MC Steps during a complete oscillation between the two chambers. The total active force $\boldsymbol{F}^{\rm tot}_{xy}$ (excluding the $z$ direction) is shown with the red arrow for each snapshot. (B) The oscillation along the $x$ coordinate of the vesicle over time. The $x_{\rm min}$, $x_{\rm mean}$, and $x_{\rm max}$ of all the vertices of the vesicle are shown in blue, grey, and red solid lines respectively. Black dashed lines indicate the dimensions of the dumbbell shaped-confinement, where, $L_{x1}=36$, $L_{x2}=78$, $L_{y1}=16$, and $L_{y2}=32$ in the units of $l_{\rm min}$. The time evolution of the $x$ component of the total force $F^{\rm tot}_{x}$ is shown in orange (scale on the right). Green vertical dotted lines denote the times of the snapshots in (A). (C) Different protein clusters are shown in different colours on a $x-y$ plane from the top view during the polarity flip. We indicate the actin flow contribution from each cluster in the corresponding coloured arrow at the mean position of that cluster. The total actin flow is indicated using a black arrow at the centre of mass of the vesicle. We use the adhesion strength $E_{ad}=3k_BT$, active force parameter $F=3k_BT l_{\rm min}^{-1}$, and the coupling parameter $\beta=20~D/k_BT$. (D) On the left, a schematic diagram of the vesicle where the curved proteins are coloured with red if it's $x$ coordinate $x_i>x_{\rm CM}$, and with blue if $x_i<x_{\rm CM}$. The centre of mass is denoted by a lime-coloured marker. On the right, the mean curvature for the curved proteins for $x_i>x_{\rm CM}$ and $x_i<x_{\rm CM}$ in red and blue respectively. The black dashed lines denote the times of the force reversals along the $x$ direction.}\label{fig:box_confinement}
\end{figure}
Comparing to the hard-wall case, we find that even without the UCSP mechanism ($\beta=0~D/k_BT$) the motile vesicle maintains its polarity when hitting the soft wall (Fig.\ref{fig:soft_vs_hard_wall}E). This is facilitated by the membrane maintaining its high curvature along its leading-edge, thereby preventing the CMC cluster from breaking up into the two-arc configuration (See Movie~S-11). After spending some time stuck against the wall, the vesicle spontaneously rotates and migrates away (with significantly diminished polarization, Fig.~S-5B, C). In the presence of UCSP, the cells get stuck penetrating the barrier for longer times as the coupling strength increases (Fig.\ref{fig:soft_vs_hard_wall}F), before migrating away (See Movie~S-12).

A more extreme scattering configuration is presented in Fig.\ref{fig:tip}. Motivated by experiments we let our motile vesicle hit the triangular tip of a square-shaped barrier, edge on (the shape of the triangular tip is explained in SI section G.1). As expected, in the absence of UCSP, the vesicle loses its polarity to the immotile (See Movie~S-13) two-arc shape (Fig.\ref{fig:tip} A). As the values of $\beta$ increase the UCSP mechanism which allows the vesicle to recover its polarized shape following the scattering with the barrier, and continue its migration (See MovieS-14). In Fig.\ref{fig:tip}D we show similar scattering events in experiments using \textit{D. discoideum} cells hitting PDMS barriers(See Movie~S-15). As in the model, the cells can lose their polarity, form two competing protrusions with very large shape elongation (similar to the two-arc shape), and recover their polarity.

The most elaborate test of our vesicle's motility under confinement is shown in Fig.\ref{fig:box_confinement}A. Here, we consider a dumbbell-shaped region surrounded by rigid barriers. The configuration of this confinement is motivated by experiments that have demonstrated spontaneous cellular oscillations within this system \cite{gegenfurtner2018micropatterning,fink2020area}. In these experiments it was found that cells spontaneously oscillate along the dumbbell pattern, and our model displays very similar behaviour (Fig.\ref{fig:box_confinement}A,B). Within our model, the origin of this oscillatory behaviour is explained in Fig.\ref{fig:box_confinement}C: As the vesicle moves to the right, its leading edge reaches the rigid walls of the confining barrier, where the leading edge loses its high curvature (Fig.\ref{fig:box_confinement}D), and the leading edge cluster breaks into two (or more) clusters on either side, protruding mainly along the perpendicular directions. The contributions of these clusters to the global retrograde actin flow approximately cancel each other, and the global flow along the long axis of the pattern ($x$-axis) becomes dominated by the trailing edge CMC cluster and switches direction. As the flow direction switches, so does the polarity cue gradient and the activity of the CMC becomes strong in the new leading edge and weak at the new trailing edge, and the cell moves towards the opposite end of the dumbbell, and so on (See Movie~S-16).

In the SI section G.2 (Fig.~S-9) we demonstrate these oscillations in a dumbbell pattern confined by adhesion (See Movie~S-17), rather than rigid barriers, where the mechanism for the oscillations is identical: since the leading edge does not easily extend over the non-adhesive region, it loses its sharp edge, flattens and breaks up. We also demonstrate these oscillations in a simpler rectangular adhesive confinement (SI section H, See Movie~S-18). In this case, the cell can get stuck for longer times at the end of the rectangle, as there is less space available for the leading edge to quickly break up into two opposing parts compared to the dumbbell pattern. Indeed, cells inside confining adhesion patterns often exhibit random oscillatory dynamics in experiments \cite{fink2020area,kalukula2024actin}, sometimes getting stuck at the ends of the patterns before switching their direction of migration \cite{zhou2020quasi}. 

Our model is based on a physical mechanism that inhibits the leading edge at the pattern's edge, namely its curvature sensitivity, which explains both the behavior for adhesive and topographic confinement. Additional biochemical feedbacks may also contribute \cite{camley2013periodic, Zadeh2024arXiv}. Our results are also relevant to cellular oscillations observed when cells form their own adhesive ``confinement" by deposition of extracellular matrix (ECM) \cite{d2021cell,perez2024deposited}.




\section{Discussion and conclusion}

Recently we have demonstrated that the coupling of curvature (through CMC) and recruitment of active protrusive forces due to actin polymerization, together with surface adhesion \cite{sadhu2021modelling}, provides a powerful organizing principle that can explain a variety of cellular shape dynamics and migration patterns  \cite{sadhu2023minimal,sadhukhan2023modelling,sadhu2024minimal}. Here we extended this model by implementing a simplified mechanism that couples the membrane organization of the CMC to an internal net actin flow that induces a polarity cue gradient across the cell. This polarity cue in turn introduces long-range inhibition of the local forces exerted by the CMC, thereby completing the feedback between the CMC organization and global polarization of the cell \cite{maiuri2015actin,ron2020one}. 

This extension greatly increases the robustness of the polarized vesicle in our model. It allows us to use our model to explain a large variety of cellular dynamics which are observed in living cells, such as the relation between cell speed and aspect-ratio, cell-barrier scattering, and cellular oscillations in different types of geometric confinements. The agreement between the experiments and the model emphasizes that curved protein complexes are crucial in the formation and dynamics of lamellipodia-driven cell migration \cite{wu2024wave}, and explain the sensitivity of the lamellipodium's stability to its leading-edge curvature \cite{jiang2023switch}. We demonstrate that simple, and therefore general (not cell-type specific), physics-based mechanisms play essential roles in directing cellular shape and migration. Biological and biochemical complexity allows cells to exert more precise control over these physical mechanisms, in response to different external conditions.\\

\section{ACKNOWLEDGMENTS} {N.S.G. is the incumbent of the Lee and William Abramowitz Professorial Chair of Biophysics, and acknowledges support by the Israel Science Foundation (Grant No. 207/22). This research is made possible in part by the historic generosity of the Harold Perlman Family. The research of CB and C.M.-T. has been partially funded by the Deutsche Forschungsgemeinschaft (DFG), Project-ID No. 318763901–SFB1294 \\} 

\textbf{Author contributions}---{SS and NSG designed the computational research; SS implemented the code and ran the simulations and data analysis; CMT and CB designed the experimental research; CMT performed the experiments; SP and AI developed the original code; SS and NSG wrote the paper, edited by CMT, CB, AI. All authors contributed to the article and approved the submitted version.}\\
\textbf{Author declaration}---{The authors declare no competing interest.}

\bibliography{shubhadeep_biophys}

\end{document}


\newcommand{\be}{\begin{equation}}
\newcommand{\ee}{\end{equation}}
\newcommand{\bea}{\begin{eqnarray}}
\newcommand{\eea}{\end{eqnarray}}
\newcommand{\nn}{\nonumber}

\title{Supplementary material: Modelling how lamellipodia-driven cells maintain persistent migration and interact with external barriers}

\author{Shubhadeep Sadhukhan}
\email{shubhadeep.sadhukhan@weizmann.ac.il}
\affiliation{%
 Department of Chemical and Biological Physics, Weizmann Institute of Science, Rehovot, Israel
}

\author{Cristina Martinez-Torres}
\email{martineztorres@uni-potsdam.de}
\affiliation{Institute of Physics and Astronomy, University of Potsdam, Potsdam 14476, Germany}
\author{Samo Peni\v{c}}
\email{samo.penic@fe.uni-lj.si}
\affiliation{%
Laboratory of Physics, Faculty of Electrical Engineering, University of Ljubljana, Ljubljana, Slovenia
}
\author{Carsten Beta}
\email{beta@uni-potsdam.de}
\affiliation{Institute of Physics and Astronomy, University of Potsdam, Potsdam 14476, Germany}
\affiliation{Nano Life Science Institute (WPI-NanoLSI), Kanazawa University, Kanazawa 920-1192, Japan}

\author{Ale\v{s} Igli\v{c}}
\email{ales.iglic@fe.uni-lj.si}
\affiliation{%
Laboratory of Physics, Faculty of Electrical Engineering, University of Ljubljana, Ljubljana, Slovenia
}
\affiliation{%
Laboratory of Clinical Biophysics, Faculty of Medicine, University of Ljubljana, Ljubljana, Slovenia
}
\author{Nir Gov}%
\email{nir.gov@weizmann.ac.il}
\affiliation{%
 Department of Chemical and Biological Physics, Weizmann Institute of Science, Rehovot, Israel
}
\date{\today}
\maketitle
\renewcommand{\thefigure}{S-\arabic{figure}}
\renewcommand{\thesection}{S-\arabic{section}}
\renewcommand{\theequation}{S-\arabic{equation}}

\section*{Materials and methods}
\subsection{Theoretical model}\label{SI_sec:model}
We modelled the cell membrane as a three-dimensional vesicle which is described by a closed surface with $N$ vertices connected to its neighbours by bonds and it forms a dynamically triangulated, self-avoiding network, with the topology of sphere~\cite{sadhu2021modelling,sadhu2023minimal}.  $\boldsymbol{r_i}$ is the position vector of the $i$th vertex. All the lengths are measured in a scale of $l_{\rm min}$.
There is a percentage $\rho=100N_c/N$ of vertex sites that represent the proteins that induce cytoskeletal active forces. The vesicle is placed on a uniform adhesive substrate. The vesicle energy has four components: The bending energy is given by,
\begin{equation}
    W_b=\frac{\kappa}{2}\int_A (C_1+C_2-C_0)^2~dA,
\end{equation}
where, $\kappa$ is the bending rigidity, $C_1$, $C_2$ are the principal curvatures and $C_0$ is the spontaneous curvature.  We consider the spontaneous curvature  $C_0=1 l_{\rm min}^{-1}$  for the curved protein sites represented in red and blue represents the bare membrane for which $C_0=0$. We set $\kappa=20 k_B T$ throughout the paper except for the cases we mentioned. The protein-protein interaction energy is given by,
\begin{equation}
    W_d=-w\sum_{i<j}\mathcal{H}(r_0-r_{ij})
\end{equation}
where $\mathcal{H}$ is the Heaviside function, $r_{ij}=|\boldsymbol{r_i}-\boldsymbol{r_j}|$ is the distance between protein sites, $r_0$ is the range interaction and $w$ is the strength. We set the parameter $w$ to 1$k_BT$ throughout the paper. The energy due to the active force is given by,
\begin{equation}
    W_F=-F\sum_i \hat{n_i}\cdot \boldsymbol{r_i}
    \label{eq:active_energy}
\end{equation}
where $F$ is the magnitude of the active force, $\hat{n_i}$ is the outward unit normal vector of the $i$th protein site vertex and $\boldsymbol{r_i}$ is the position vector of the protein. Finally, the adhesion energy due to the extra-cellular substrate is given by,
\begin{equation}
    W_A=-\sum_{i'} E_{\rm ad}
\end{equation}
where $E_{\rm ad}$ is the adhesion strength, and the sum runs over all the adhered vertices to the substrate i.e., the $z$ component of the vertex position is within a range $z_{\rm ad}<z_i<z_{\rm ad}+\delta z$. We set $\delta z=1l_{\rm min}$ throughout the paper.

\subsection{Calculation of direction of net actin flow}\label{SI_sec:DIR}
For simplicity, we want to map the whole three-dimensional problem into a simplified one-dimensional model as the concentration profile of the inhibiting molecules is calculated for one-dimensional case~\cite{ron2020one}. We want to approximate the direction along which we want to calculate the variation of the concentration of inhibitors. First, curved-membrane protein (CMP) components on the vesicle membrane are clusterized. Let us consider the $i$th cluster that consists of $N_i$ number of proteins. The positions of the $j$th protein-vertex of the $i$th cluster is ($x^i_j,~y^i_j,~z^i_j$).  We approximated the actin flow from the protein site towards the centre of mass of the vesicle $\boldsymbol{r}_{\rm CM}$($x_{\rm CM},~y_{\rm CM},~z_{\rm CM}$) and the magnitude of the flow is proportional to the active force. We compute the centre of mass of the vesicle by taking the average position of all vertices on the membrane. Therefore, we find a vector $\boldsymbol{c}^i$=($c^i_x, ~c^i_y, ~c^i_z$) for the $i$th cluster by,
\begin{eqnarray}
c^i_x=\frac{1}{N_i}\sum_{j} F^i_j (x^i_j-x_{\rm CM})\nonumber\\
c^i_y=\frac{1}{N_i}\sum_{j} F^i_j (y^i_j-y_{\rm CM})\nonumber\\
c^i_z=\frac{1}{N_i}\sum_{j} F^i_j (z^i_j-z_{\rm CM})
\end{eqnarray}
where, $j$ runs over all the vertices within the $i$th cluster. $F^i_j$ is the magnitude of the active force by the $j$th vertex in the $i$th cluster.
For the direction of net actin flow, we fit a three-dimensional line through the centre of mass and the $\boldsymbol{c}^i$ vectors for all the clusters with a weightage proportional to their size $N_i$.

Next, we find the symmetric covariance matrix as follows,
\begin{equation}
    C=\begin{bmatrix}
    a_{xx}& a_{xy} & a_{xz}\\
    a_{yx} & a_{yy} & a_{yz}\\
    a_{zx} & a_{zy} & a_{zz}
    \end{bmatrix}.
\end{equation}
where the elements of this matrix are given by,
\begin{eqnarray}
    a_{xx}=\sum_i{N_i (c^i_x-x_{\rm CM})^2}\nonumber\\
    a_{yy}=\sum_i{N_i (c^i_y-y_{\rm CM})^2}\nonumber\\
    a_{zz}=\sum_i{N_i (c^i_z-z_{\rm CM})^2}\nonumber\\
    a_{xy}=a_{yx}=\sum_i{N_i (c^i_x-x_{\rm CM})(c^i_y-y_{\rm CM})}\nonumber\\
    a_{yz}=a_{zy}=\sum_i{N_i (c^i_y-y_{\rm CM})(c^i_z-z_{\rm CM})}\nonumber\\
    a_{zx}=a_{xz}=\sum_i{N_i (c^i_z-z_{\rm CM})(c^i_x-x_{\rm CM})}.
\end{eqnarray}

Then, we find the eigenvector $\boldsymbol{\hat{e}}$ corresponding to the highest eigenvalue using the ``power method'' with a tolerance of $10^{-3}$. This method gives the direction of net actin flow with an inaccuracy up to a sign reversal. To correct this we use a physical condition that the total active force and the net actin flow are opposite to each other, i.e., $\boldsymbol{F^{\rm tot}}\cdot \boldsymbol{\hat{e}}<0$.
\subsection{UCSP time step}
First, we find the axis of the net actin flow using the cluster information. Now, we find the projection of each vertex position with respect to the centre of mass of the vesicle on the computed flow axis. This projection for the $i$th vertex is given by,
\begin{equation}
    p_i=-(\boldsymbol{r}_i-\boldsymbol{r}_{\rm CM})\cdot\boldsymbol{\hat{e}}.
    \label{eq:proejction}
\end{equation}
The negative sign is for the convention so the maximum projection $p_{\rm max}$ and $p_{\rm min}$ denote the front and rear of the vesicle. From this projection, we compute the concentration of the polarity cue component, and hence we get the protrusive active force $\tilde{F}$ (Eq.~3) including the inhibition effect. As the total amount of the polarity cue is constant $c_{\rm tot}$, we need to modify the factor in the concentration profile of the polarity cue given in previous work~\cite{ron2020one} for one-dimension. This factor is modified due to the fact we have a three-dimensional vesicle and therefore we need to integrate over the whole vesicle volume even though the variation of the approximated net actin flow is in one direction. Therefore, an additional factor $({p_{\rm max}-p_{\rm min}})/{V_{\rm ves}}$ get multiplied to the one-dimensional form given in \cite{ron2020one}. Here, $V_{\rm ves}$ is the volume of the vesicle and $p_{\rm max}-p_{\rm min}$ denotes the elongation of the vesicle in the direction of the net actin flow. The complete updated form is given in the Eq.~2. We repeat this polarity cue concentration calculation until the net actin flow direction converges to a given tolerance level of 0.1\%. 
If the change of direction of net actin flow due to the shape change of the vesicle is more than 10\% we repeat the calculation of the concentration of polarity cues. If $\boldsymbol{{e}}$=($e_1,~e_2,~e_3$) is the net actin flow at the step when we computed the concentration profile and $\boldsymbol{{e'}}$=($e'_1,~e'_2,~e'_3$) is the of net actin flow after the change in vesicle shape, we find the relative change in a vector by,
\begin{equation}
    {\rm error}=\frac{(e'_1-e_1)^2+(e'_2-e_2)^2+(e'_3-e_3)^2}{e_1^2+e_2^2+e_3^2}.
    \label{eq:error}
\end{equation}
If the error becomes greater than $0.1$, i.e., an error of 10\% we repeat another UCSP calculation.\\

\begin{table}[htbp]
    \centering
\begin{tabular}{
|p{5cm}||p{2cm}||p{2cm}|}
 \hline
 \multicolumn{3}{|c|}{\textbf{Important Parameters List}}\\
 \hline
 Number of vertices& $N$ &1447\\
 \hline
 Bending rigidity & $\kappa$ & 20$k_BT$\\
 \hline
 Protein-protein interaction energy& $w$ & 1$k_BT$\\
    \hline
 Intrinsic curvature of proteins & $C_0$ & 1 $l_{\rm min}^{-1}$\\
     \hline
 Range of adhesion & $\delta z$ & $1 l_{\rm min}$ \\
     \hline
$z$ coordinate of adhesion plane & $z_{\rm ad}$ & $-10.725l_{\rm min}$ \\
  \hline
 Diffusion coefficient of inhibitors & $D$ & 4000\\
  \hline
 Total amount of inhibitors & $c_{\rm tot}$ & 4000\\

    \hline
 Wall position $x$ coordinate & $w_x$ & $25l_{\rm min}$ \\
    \hline
 Wall position $y$ coordinate & $w_y$ & $0l_{\rm min}$ \\
     \hline
Spring constant of the wall & $k_{w}$ & $0.1 k_BT l_{\rm min}^{-2}$ \\

\hline
Smoothness parameter for the triangular tip & $r$ & $5 l_{\rm min}$ \\
\hline
Asymptotic angle for the triangular tip & $\theta$ & $45^o$ \\
\hline
Peak force for blowing from behind & $F_{\rm blow}$ & $20 k_BT l_{\rm min}^{-1}$ \\
     \hline
Range for blowing from behind & $\Delta x$ & $2 l_{\rm min}$ \\
\hline
\multirow{2}{*}{Rectangular shaped adhesive pattern} & $L_{x}$ & 80 $l_{\rm min}$\\
&$L_{y}$ & 22 $l_{\rm min}$\\
\hline
\multirow{4}{*}{Dumbbell shaped adhesive pattern} & $L_{x1}$ & 40 $l_{\rm min}$\\
&$L_{x2}$ & 80 $l_{\rm min}$\\
& $L_{y1}$ & 14 $l_{\rm min}$\\
&$L_{y2}$ & 32 $l_{\rm min}$\\

\hline
\multirow{4}{*}{Dumbbell shaped confinement box} & $L_{x1}$ & 36 $l_{\rm min}$\\
&$L_{x2}$ & 78 $l_{\rm min}$\\
& $L_{y1}$ & 16 $l_{\rm min}$\\
&$L_{y2}$ & 32 $l_{\rm min}$\\
 \hline

\end{tabular}
 \caption{All the important  parameters used in the simulations}
 \label{table:}
\end{table}

\begin{figure}
    \centering
    \includegraphics[scale=0.5]{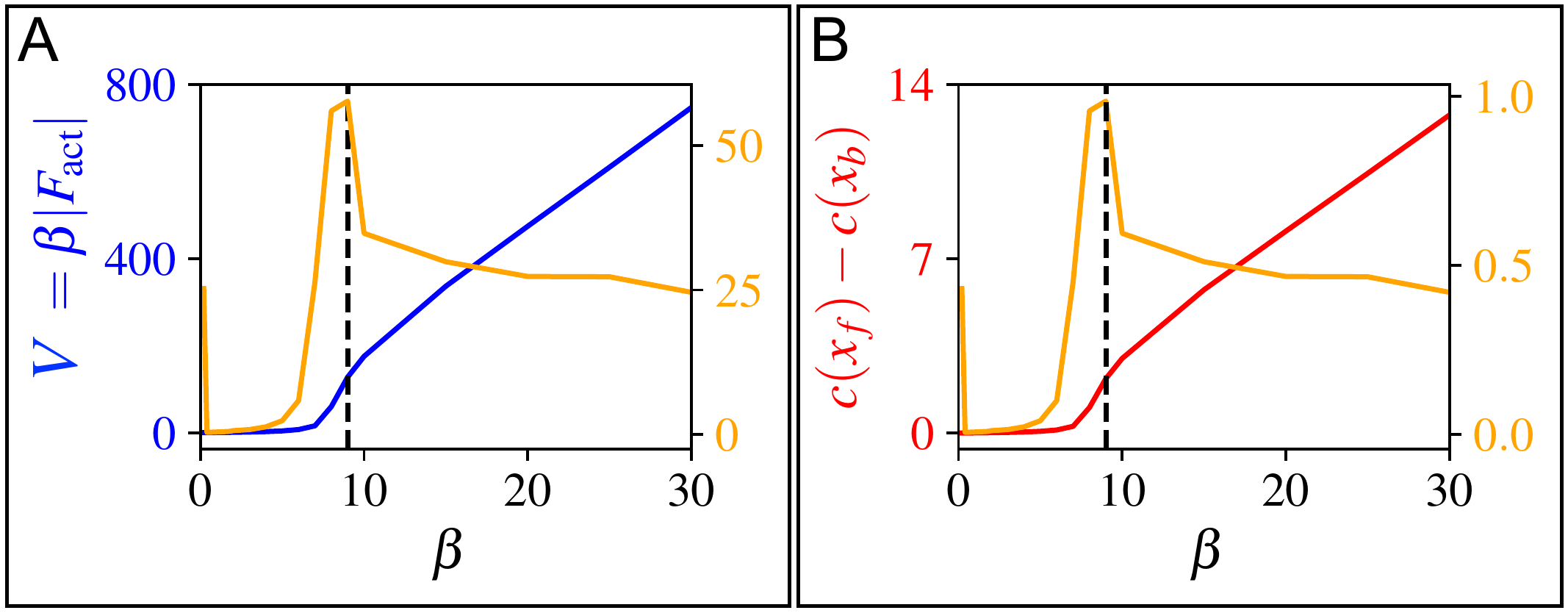}
    \caption{Demonstration of critical $\beta$: We turned off all the vertex movement and bond flips for this test. The orange lines in both panels show the first derivatives of the quantities net actin flow $V=\beta|F^{\rm tot}|$ and asymmetry in polarity cue concentration $c(p_f)-c(p_b)$ shown in the panel A and B respectively.  The derivative shows a peak at the point of the critical coupling parameter denoted by a black dashed line. We used a two-arc-shaped vesicle which was spread on the adhesion $E_{\rm ad}=1~k_BT$ and $F=3 k_BT~l_{\rm min}^{-1}$. }
    \label{fig:critical_beta}
\end{figure}
\begin{figure}
    \centering
    \includegraphics[scale=0.45]{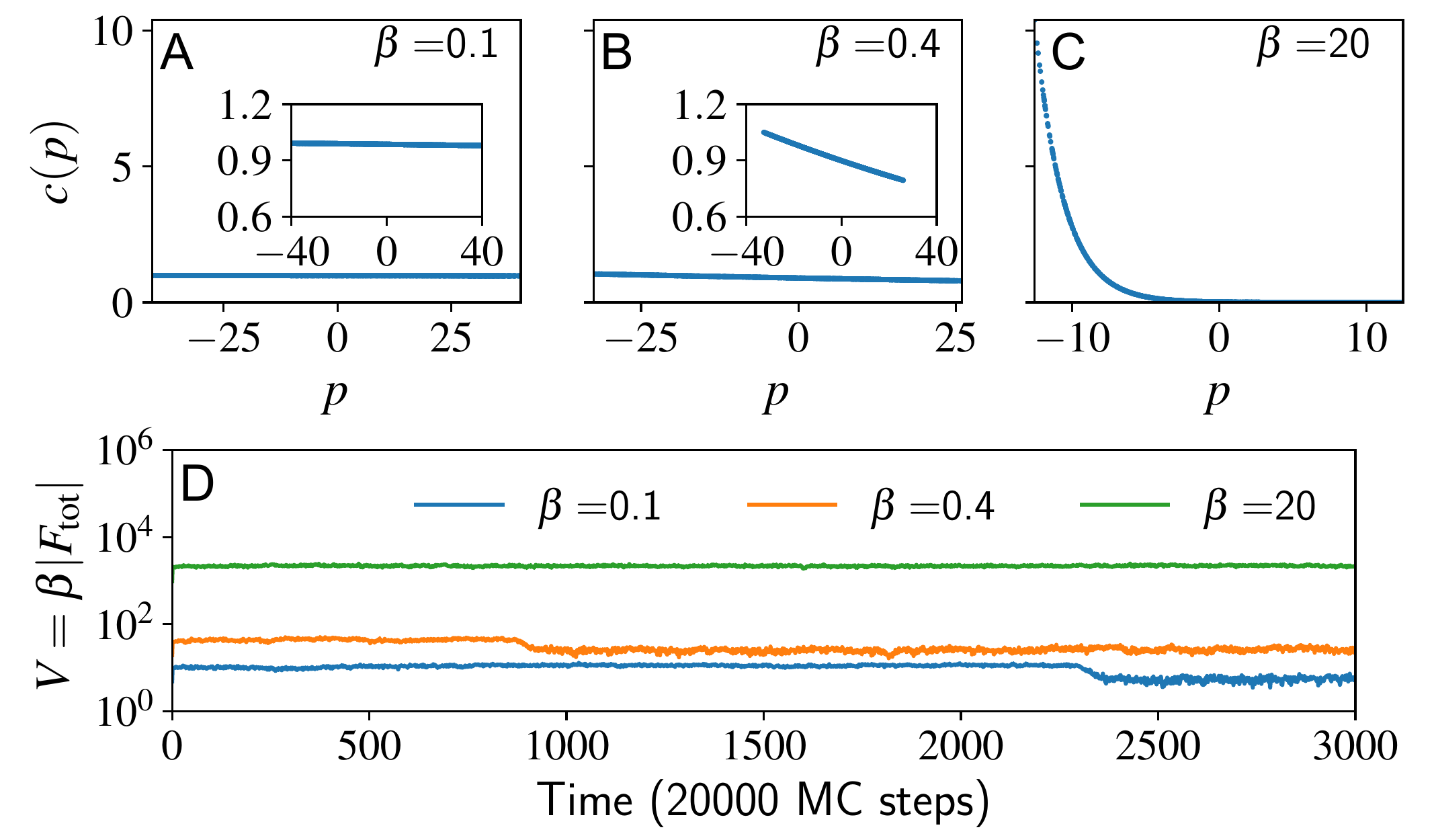}
    \caption{We showed the concentration profile along the UCSP axis at the final time 2999 in the units of 20000 MC steps for the cases $\beta=0.1, ~0.4,$ and $20$ in units of $D/k_BT$ respectively in (A-C). We can look into the concentration profile in a zoomed inset for the cases of $\beta=0.1,~0.4~D/k_BT$. We showed the time series of the magnitude of internal actin flow $V$ in D for different non-zero $\beta$.}
    \label{fig:breaking_SI}
\end{figure}
\subsection{Breaking of protein cluster}\label{sec:breaking}
We demonstrated that as the coupling between the asymmetry in polarity cues and the actin flow ($\beta$) increases, the UCSP mechanism can stabilize the crescent-shaped vesicle and hinder the transition to the non-motile shape (Fig.~2).
We showed the concentration profile of the polarity cues over the ``mini-cell'' (vesicle) along the net actin flow axis for various coupling strength $\beta=0.1, ~0.4,$ and $20$ in units of $D/k_BT$ as shown in Fig.~\ref{fig:breaking_SI} (A-C). When, $\beta$ is too small the concentration of such polarity cues is nearly constant. However, for the high value of $\beta$ can create a exponential fall of the concentration of polarity cues at the front of the  ``mini-cell''. Next, we can see that time-series plot of the net actin flow $V=\beta|F^{\rm tot}|$ and it increases as coupling parameter $\beta$ increases as shown in Fig.~\ref{fig:breaking_SI}D.


\subsection{Model of blowing force from behind}\label{SI_sec:blow_behind}

We model the force due to flowing fluid from the back of the vesicle. The force has the peak magnitude of $F_{\rm blow}$. It is acting on the vesicle from behind in the direction of $\hat{x}$. The blowing force acts only if the $x$ component of the outward normal is negative. This force magnitude also decays as a Gaussian-like function from its peak value $F_{\rm blow}$ at ($y_{\rm CM},~z_{\rm ad}$) where $z_{\rm ad}$ is the $z$ coordinate of the adhesive substrate. Therefore, mathematically we can write this force magnitude at some point $(x, y, z)$ as,
\begin{equation}
    F_{b}=F_{\rm blow} {\rm exp}\bigg(-\frac{1}{\sigma} \big[ (y-y_{\rm CM})^2+(z-z_{\rm ad})^2\big]\bigg),
    \label{eq:blow_force}
\end{equation}
where $\sigma=100 l_{\rm min}^2$.
\begin{figure}
    \centering
    \includegraphics[scale=0.33]{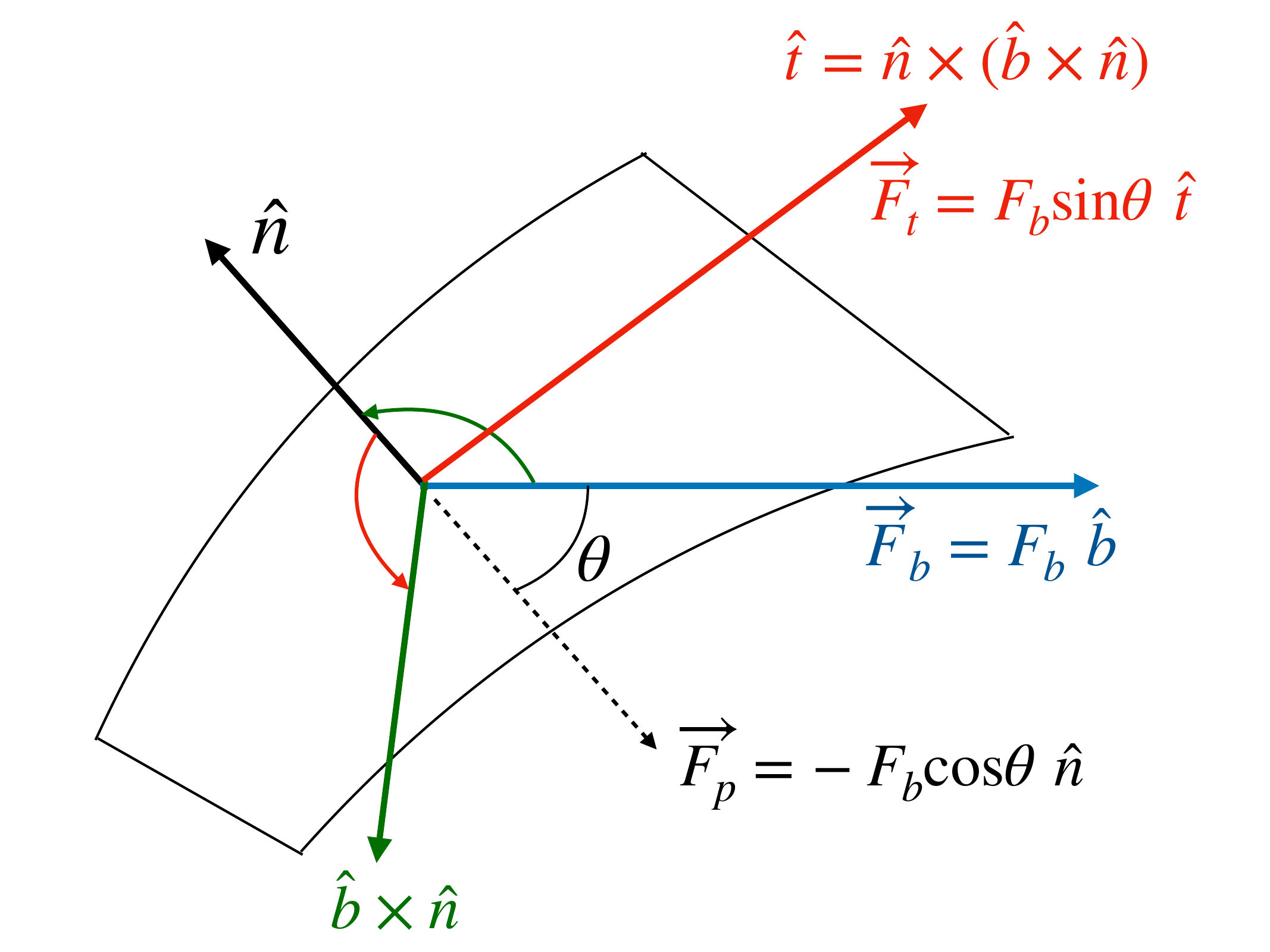}
    \caption{This force has two components $\boldsymbol{F_p}=-F_b \cos\theta~\hat{n}$  and $\boldsymbol{F_t}=F_b \sin\theta~\hat{t}$. The first one is the force due to pressure in the opposite direction of the outward normal of the membrane. The next one is the tangential drag which is tangential to the membrane. The figure shows the calculation of the directions.}
    \label{fig:blow1}
\end{figure}
Note that the blowing force is not varying with $x$ but it acts only from the back within a range from $x_{\rm back}$ to $x_{\rm back}+\Delta x$, where, $x_{\rm back}$ is the $x$ coordinate of the vesicle at the back and $\Delta x=2 l_{\rm min}$ is the range of the force. Now, this blowing force $F_{b}$ is decomposed into two parts, (a) Pressure part: $\boldsymbol{F_p}=-F_b \cos\theta~\hat{n}$, (b) Tangential drag force: $\boldsymbol{F_t}=F_b \sin\theta~\hat{t}$  where, $\hat{n}$ is the unit outward normal and $\hat{t}$ is the direction of tangential drag due to the blowing force from behind as shown in Fig.~\ref{fig:blow1}.

\subsection{Wall implementation}\label{SI_sec:wall_implementation}
\begin{figure}
    \centering
    \includegraphics[scale=0.24]{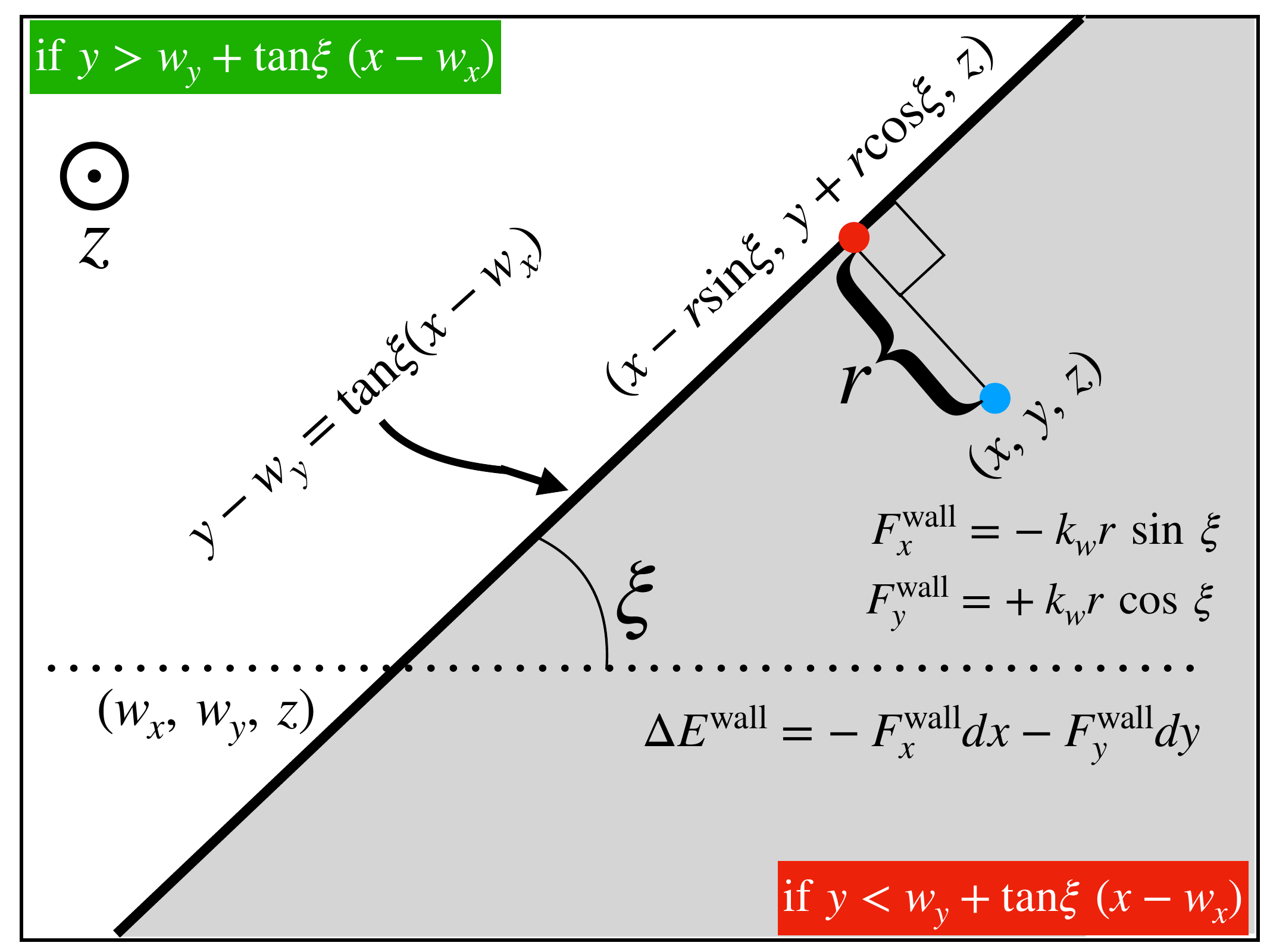}
    \caption{Schematic diagram of the wall implementation: The wall is specified by a straight line on $x-y$ plane which passes through the point ($w_x, ~w_y$) and makes an angle $\xi$ with the $x$ axis.  The wall is extended semi-infinitely in $z$ direction (Outward normal of the page) for the region $z\geq z_{\rm ad}$. We showed the region beyond the wall using a grey shade. Mathematically, if the condition $y<w_x+{\rm tan} ~\xi(x-w_x)$ holds, the vertex is trying to penetrate the wall. For a completely rigid wall ($k_w\rightarrow \infty$), the energy cost is infinite. Hence, no penetration is allowed. In the case of a soft wall, we model the softness of the wall by Hook's law where the vertex of the vesicle feels a restoring force depending on the position where the vertex is trying to penetrate. If ($x,~y,~z$) is the position of the vertex, then the restoring force applied on it is given by, ($-k_wr~{\rm sin }\xi,~k_wr~{\rm cos}\xi,~0$), where $r=(x-w_x){\rm sin}\xi-(y-w_x){\rm cos \xi}$ is the perpendicular distance of the wall from the vertex when the vertex is in the grey shaded region. Therefore, if the vertex displaced $\boldsymbol{dr}=(dx,~dy,~dz)$ it costs an amount of energy due to the soft wall is $\Delta E^{\rm wall}=-F^{\rm wall}_xdx-F^{\rm wall}_y dy$.}
    \label{fig:wall_implementation}
\end{figure}
\begin{figure}
    \centering
    \includegraphics[width=\linewidth]{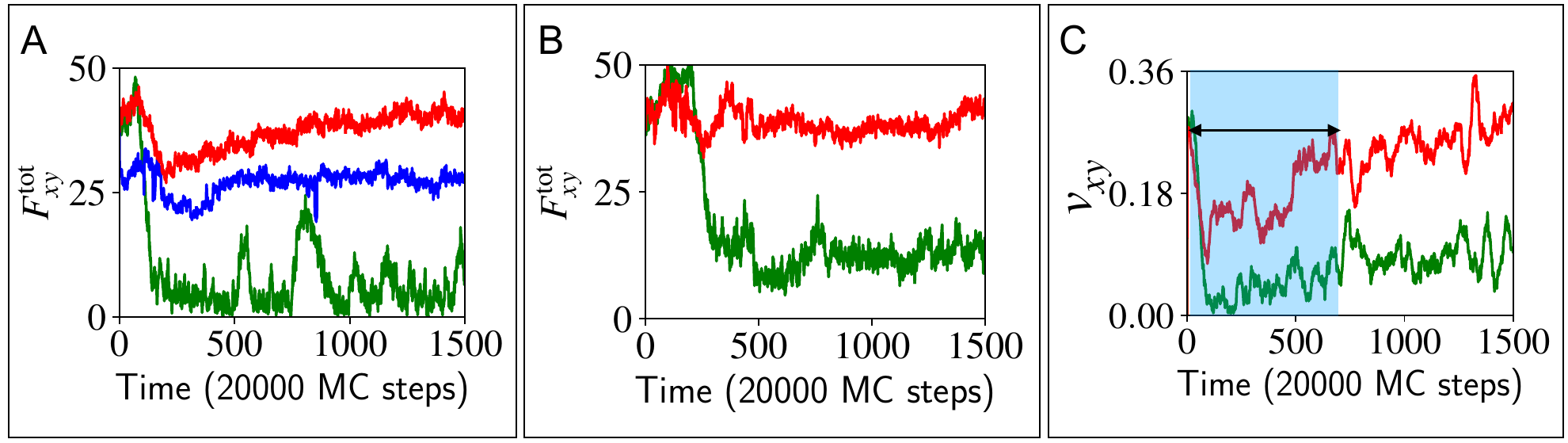}
    \caption{Total planar force $F_{xy}^{\rm tot}$ from the simulation is shown in (A) and (B) for rigid and soft barriers respectively. We denoted the case of $\beta=0,~6,$ and $15$ in units of $D/k_BT$ with the green, blue and red colours respectively. (C) The planar velocity of the vesicle during its interaction with a soft barrier for the cases $\beta=0~D/k_BT$ and $15~D/k_BT$ in green and red colours respectively. It shows a similar dip in speed (denoted by a blue-shaded rectangle).}
    \label{fig:SI_wall}
\end{figure}

In our model, the vesicle is adhered to a plane $z=z_{\rm ad}$. Therefore, it moves on the $x-y$ plane. We implement the wall by drawing a line on $x-y$ plane and the wall is semi-infinitely extended in $z$ direction for the region $z\geq z_{\rm ad}$. Mathematically, we can specify the wall uniquely by telling the coordinates ($w_x,~w_y$) through which the line passes and the angle $\xi$ with the $x$ axis as shown in Fig.~\ref{fig:wall_implementation}. The equation of this line is given by,
\begin{equation}
    y=w_y+{\rm tan} \xi (x-w_x).
    \label{eq:wall}
\end{equation}
Let us consider a vertex move that tries to place a vertex at the coordinates ($x, ~y, ~z$). Then, we detect the vertex is trying to penetrate the wall by the condition if $y<w_y+{\rm tan} \xi (x-w_x)$. The rigidity of a wall is characterized by a spring constant $k_w$ corresponding to the wall. Higher $k_w$ implies a more rigid wall, whereas smaller $k_w$ implies a softer wall. If the wall is completely rigid, i.e., $k_w\rightarrow\infty$ it costs infinite energy. Therefore, this vertex move is abandoned. 

\begin{figure}
    \centering
    \includegraphics[width=1\linewidth]{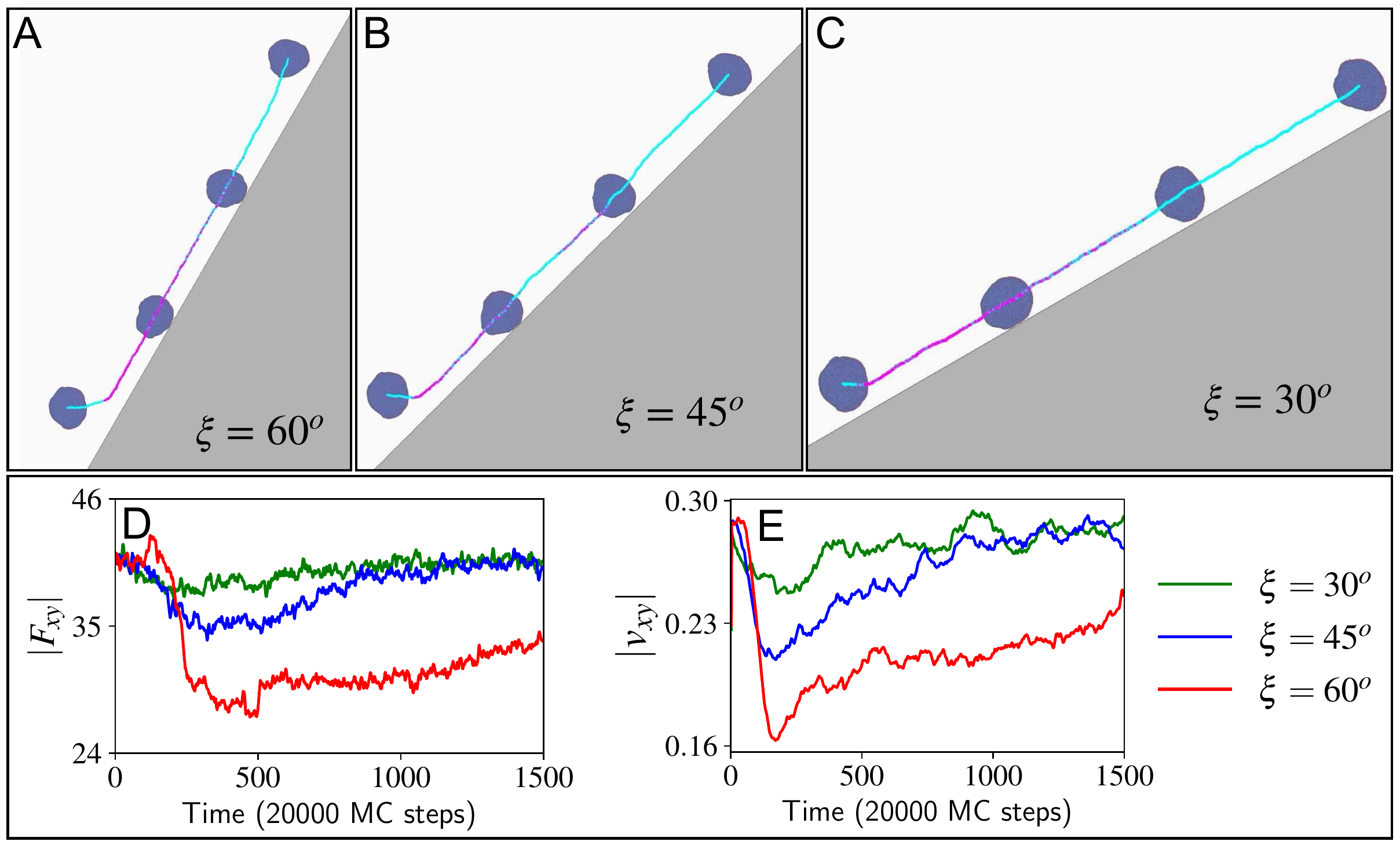}
    \caption{Scattering of the vesicle with slanted rigid walls ($\xi\neq 90^o$): We showed the trajectory of the vesicle scatters from the walls specified by the angles $\xi=60^o,~45^o,~30^o$ through the point $(50,~0)$ are shown in (a)-(c) respectively. Used parameter $E_{\rm ad}=3k_BT,~F=2k_BTl^{-1}_{\rm min}$, $\beta=15~D/k_BT$ . The magenta colour on the trajectory denotes that the vesicle touches the wall, whereas the cyan colour on the trajectory denotes that the vesicle does not touch the wall. (d)-(e) The planar force magnitude $|F_{xy}|$ and planar velocity magnitude $|v_{xy}|$ (both averaged over 5 different realizations) are shown for $\xi=30^o,~45^o,$ and $60^o$ with green, blue and red respectively. It shows the loss of polarity before repolarization increases as the inclination angle $\xi$ increases.}
    \label{fig:slanted_wall}
\end{figure}

Besides the completely rigid walls, the wall can also be a soft one, i.e., $k_w$ is finite. Then, the energy cost for such a vertex move will be finite. Smaller $k_w$ means a softer wall. If a vertex tries to penetrate the wall (i.e., in the shaded region as shown in Fig.~\ref{fig:wall_implementation}), then the vertex feels a restoring force following Hook's law. We find the perpendicular from the vertex position (Blue dot in Fig.~\ref{fig:wall_implementation}) to the wall. Hence the soft wall will try to push back the vertex by applying a restoring force that is directed normally to the equilibrium position (red dot in Fig.~\ref{fig:wall_implementation}) of the displaced wall. The coordinate of the equilibrium position is given by, $(x - r \sin\xi, y+r\cos\xi)$, where $r=(x-w_x)\sin \xi-(y-w_y)\cos\xi$ is the perpendicular distance of the wall from the vertex when the vertex is in the grey shaded region. Therefore, the restoring force felt by the vertex is given by,
\begin{eqnarray}
    F^{\rm wall}_x=-k_w r \sin \xi\nonumber &\\
    F^{\rm wall}_y=+k_w r \cos \xi.
    \label{eq:wall_force}
\end{eqnarray}
Therefore, the energy cost due to the soft wall is given by, 
\begin{equation}
    \Delta E^{\rm wall}=-F^{\rm wall} dx-F^{\rm wall}_y dy.
\end{equation}
In the main text, we already discussed the results of the vesicle's head-on collision with the rigid and soft wall barriers ($\xi=90^o$) as shown in the Fig.~5.
We placed the motile vesicle in front of a slanted rigid barriers $\xi\neq90^o$. We set the wall parameters $w_x=50 l_{\rm min}$, $w_y=0 l_{\rm min}$. The typical trajectories with the angle $\xi=30^o,~45^o,$ and $60^o$ are shown in Fig.~\ref{fig:slanted_wall}(A-C) respectively. We set the coupling $\beta$ to 15. We see the polarity loss and repolarization through the net planar force magnitude $|F_{xy}|$ and the velocity magnitude $|v_{xy}|$ in Fig.~\ref{fig:slanted_wall}(D-E). As the angle $\xi$ increases we see the most dip in planar velocity and force increases. As the collision is more like a head-on collision it breaks the polarity more strongly. As the coupling is strong, finally they gain their polarity back.

\begin{figure}
    \centering
    \includegraphics[width=0.7\linewidth]{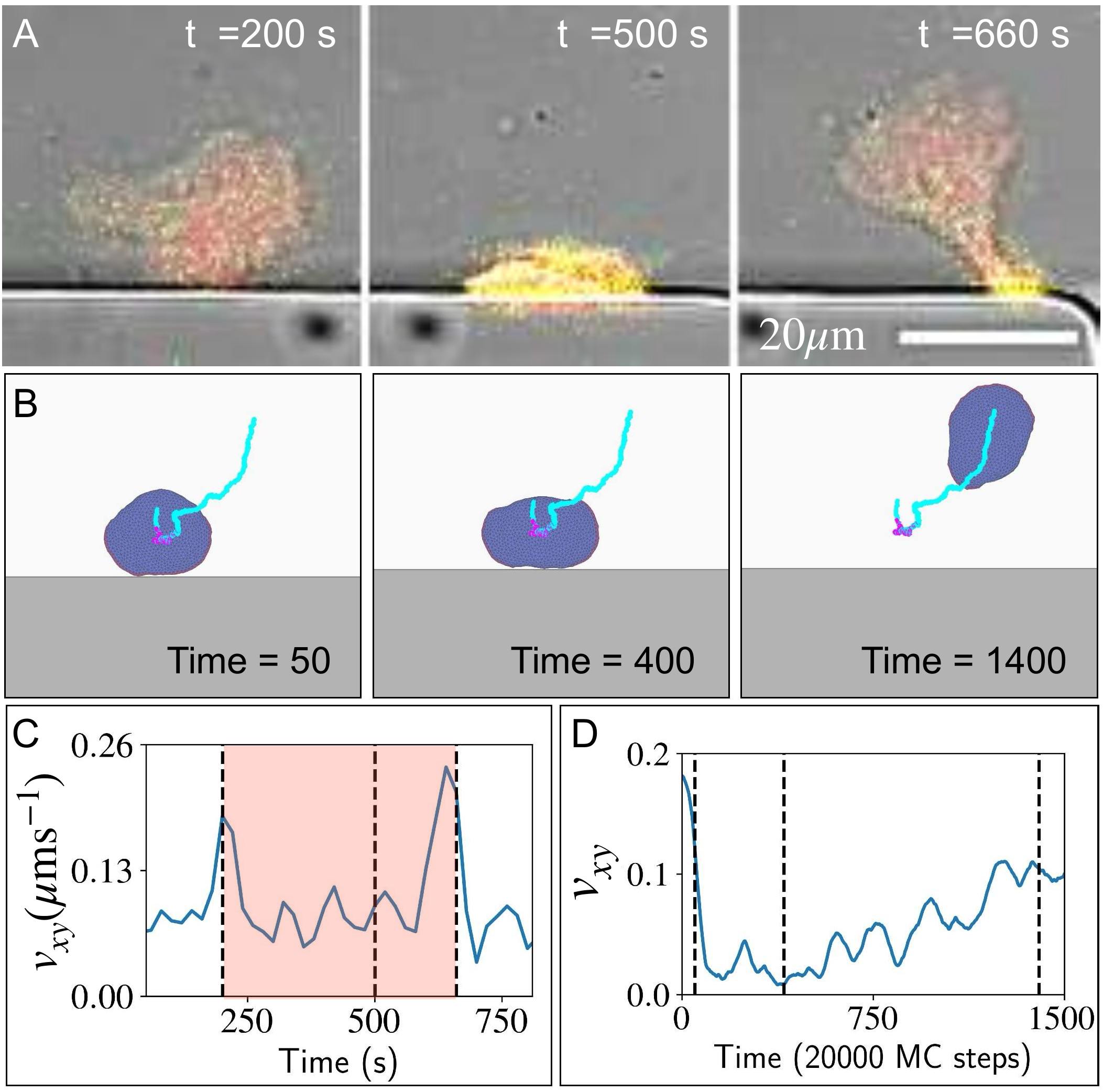}
\caption{(A) Scattering of a \textit{D. Discoideum} cell from the barrier (See Movie~\ref{mov:HW_scatter_expt})
(B) Scattering of a polar crescent-shaped vesicle from a rigid wall ($\xi=90^o,~w_x=25 l_{\rm min},~w_y=0~l_{\rm min}$) instead of hugging motion along the wall. Used parameters are $\beta=6~D/k_BT, ~E_{\rm ad}=3k_BT,~F=2k_BTl^{-1}_{\rm min}$ (See Movie~\ref{mov:HW_scatter_sim}). (C) The planar velocity $v_{xy}$ of the \textit{D. Discoideum} cell in experiments. (D) The planar velocity $v_{xy}$ of our "mini-cell" in simulations.}
    \label{fig:large_angle_scattering}
\end{figure}

\subsection{Scattering in the non-uniform geometries}
\subsubsection{Hit on a triangular tip}\label{SI_sec:triangular_tip}

We placed a crescent-shaped polar vesicle in front of a triangular-tipped geometry and let it move to hit the triangular tip.
\begin{figure}
    \centering
    \includegraphics[scale=0.2]{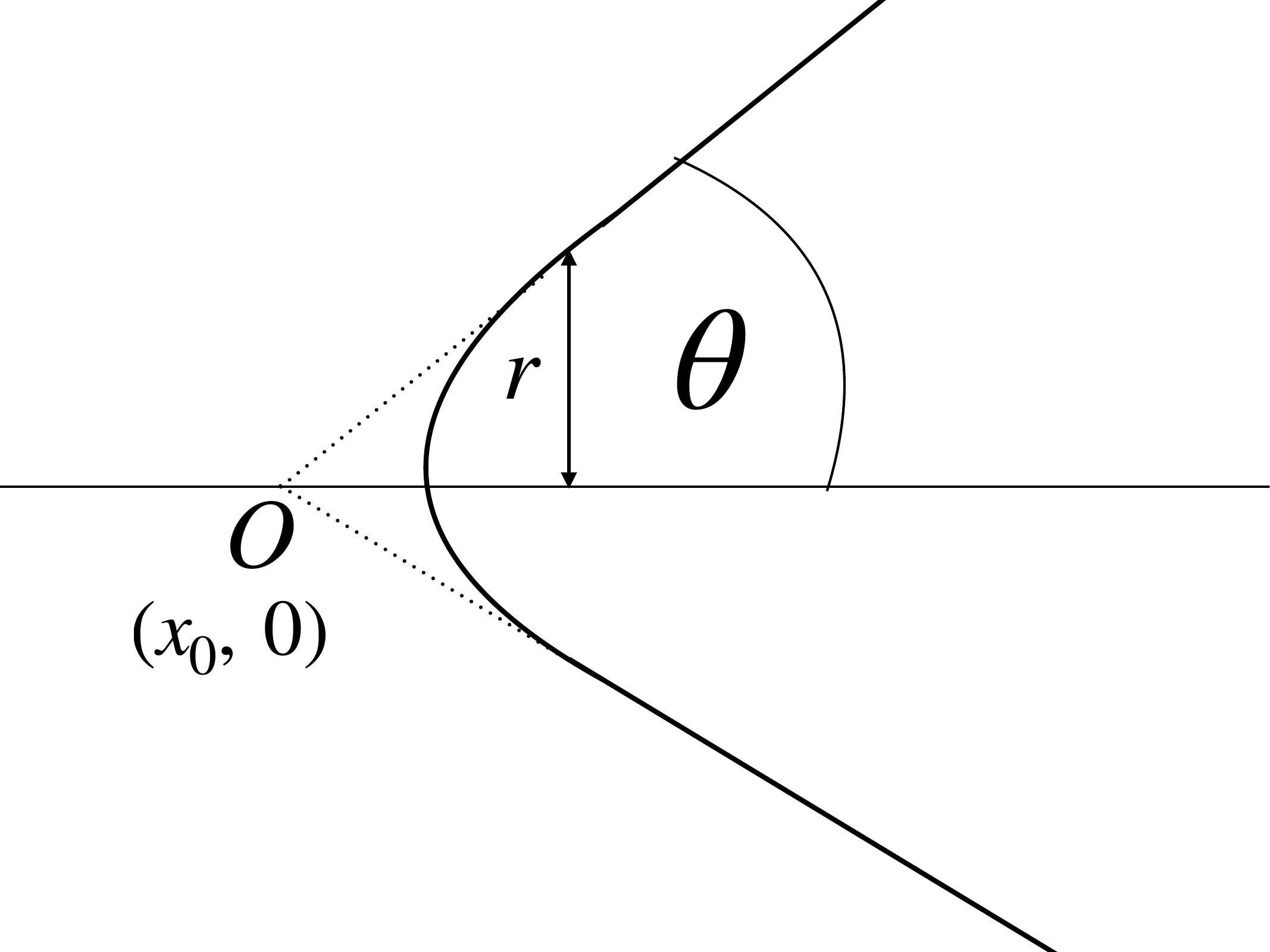}
    \caption{Schematic diagram of the rounded triangular tip used for the shape of the rigid barrier in Fig.~6.}
    \label{fig:schematic_tip}
\end{figure}
To create the triangular tip geometry we allowed the vertex movement in the correct region. The triangular tip is symmetric in $y$, about the horizontal line $y=0$. Let the sharp tipping point be located at $(x_0,~0)$ and the half angle of the triangular tip is $\theta$ as shown in Fig.~\ref{fig:schematic_tip}. If the position of the vertex of interest is ($x,~y,~z$), then we allow the vertex to move if the new position's $x$ coordinate does not exceed a maximum value $x_{\rm max}$ for a given $y$. In addition, the maximum allowed $x$ position depends on $|y|$ as it is symmetric about $y=0$. For a sharp triangular tip, the maximum allowed $x_{\rm max}$ is given by a linear function,
\begin{equation}
    x_{\rm max}=|y|/ \tan \theta+x_0.
    \label{eq:linear_tip}
\end{equation}
We showed the sharp-edged triangular tip by dotted line in Fig.~\ref{fig:schematic_tip}.

To ensure the breakage of the protein aggregate smooth, we made the tip rounded near its sharp edge. If we are very close to the tipping point, we model the tipping edge as a parabolic function in such a way it should match with the linear function at $|y|=r$. Here, $r$ is the parameter with which we define the range of the rounded edge. Now, the maximum value $x_{\rm max}$ depends on the $y$ position of the vertex and it is given by,
\begin{eqnarray}
x_{\rm max}= 
\begin{cases}
    \frac{|y|}{\tan \theta}+x_0, &  \text{if  } |y|\geq r\\
    \frac{y^2/2r+r/2}{\tan \theta}+x_0, & \text{Otherwise.} 
\end{cases}
\label{eq:rounded_tip}
\end{eqnarray}
We set $\theta=45^o$ and $r=5 l_{\rm min}$ for all the simulations done in this paper. We choose a parabolic function in such a way the piece-wise derivatives match at $|y|=r$.
We showed how a polar crescent-shaped vesicle loses its polarity and repolarizes in another direction while interacting with a triangular tip of a square-corner edge in Fig.~6.
\subsubsection{Dumbbell-shaped confinement}\label{SI_sec:dumbbell_confinement}

A ``H'' or a dumbbell-shaped confinement is implemented with rigid barriers. In doing so, we disallowed the vertex movement outside the confined region. The ``H'' shaped confinement is shown in the Fig.~7B. Let a vertex try to move to a position $(x, ~y, ~z)$. Then $x$ must satisfy the condition, $-Lx_2/2 \leq x \leq Lx_2/2$. Now, depending on $x$, we only allow those vertex moves that satisfy,
\begin{eqnarray}
&|x|\leq Lx_2/2\nonumber \\
&|y|\leq 
\begin{cases}
    Ly_1/2, & \text{if  } |x|\leq Lx_1/2\\
    Ly_2/2, & \text{if  } Lx_1/2\leq|x|\leq Lx_2/2.
\end{cases}
\label{eq:dumbbell_cond}
\end{eqnarray}
The oscillation of the vesicle within such dumbbell-shaped confinement is shown in Fig.~7A. The oscillation of $x$ position and the total active force $F^{\rm tot}_x$ is shown in  Fig.~7B. The breaking of the protein cluster and the reversal of the polarity are shown in  Fig.~7C. Fig.~7D shows the oscillation of the curvature for the curved proteins for $x_i>x_{\rm CM}$ and $x_i<x_{\rm CM}$ in red and blue respectively. When the leading edge cluster interacts with the rigid barrier of one end it gets fattened and curvature decreases. The leading edge breaks into parts and net actin flow nearly cancels. However the protein cluster in the rear end contributes more in net actin flow and protrudes strongly, hence the curvature at the rear end increases. This mechanism leads to the oscillation of the vesicle within the dumbbell-shaped rigid confinement barrier (See Movie~\ref{mov:DBLL_rigid}).  



\begin{figure}[h!]
    \centering
    \includegraphics[scale=0.33]{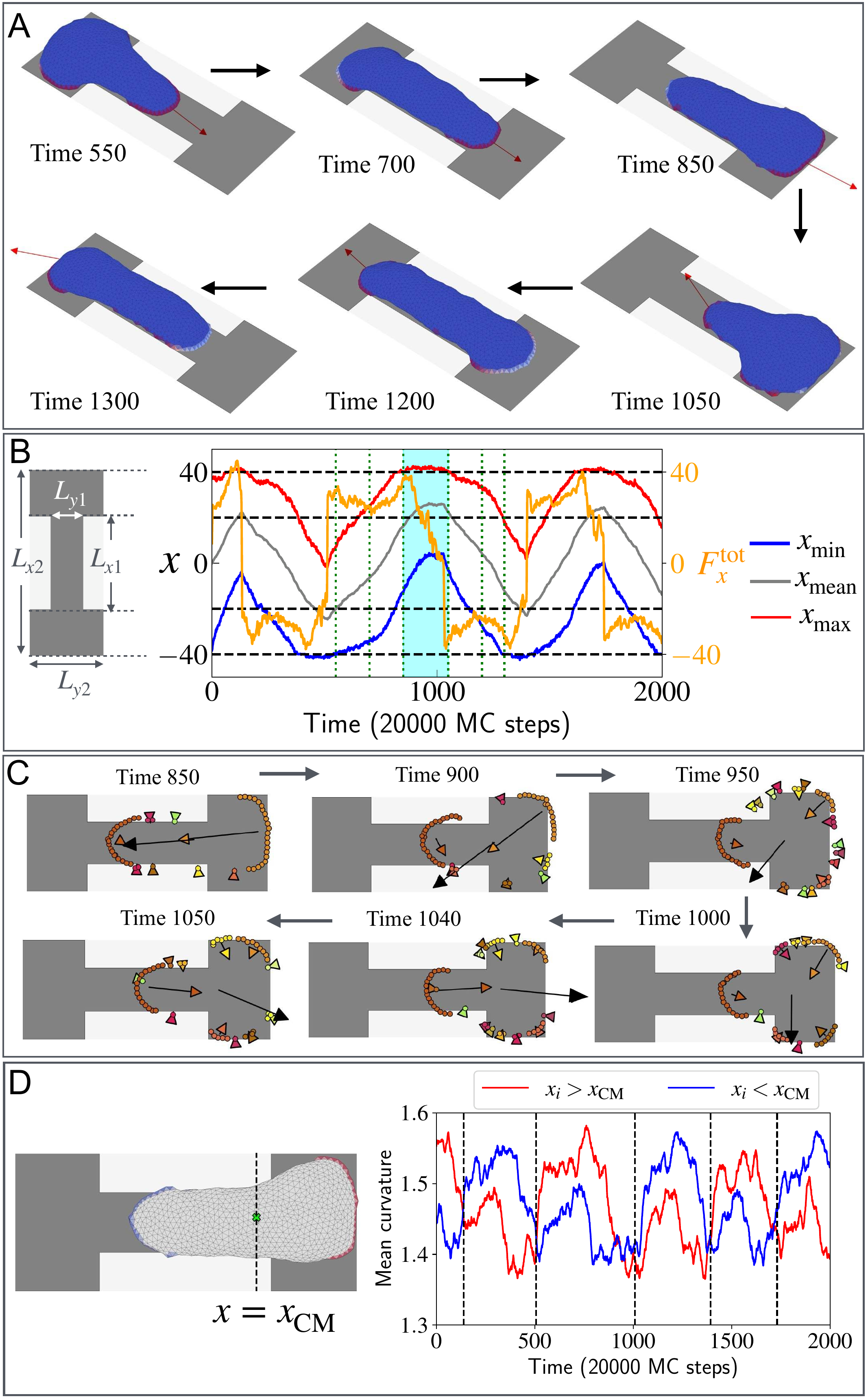}
    \caption{Spontaneous oscillations of a vesicle confined on a dumbbell-shaped adhesion patch. (A) Snapshots of the vesicle at different times, during a complete oscillation between the two chambers. The dynamics of the total planar active force $\boldsymbol{F}^{\rm tot}_{xy}$ is indicated by the red arrow. We use the adhesion strength $E_{ad}=3k_BT$, active force parameter $F=3k_BT l_{\rm min}^{-1}$, and the coupling parameter $\beta=20~D/k_BT$. (B) The oscillation of the vesicle's $x_{\rm min}$, $x_{\rm mean}$, and $x_{\rm max}$ of all the vertices (blue, grey, and red solid lines). The dimensions of the dumbbell shaped-confinement: $L_{x1}=40$, $L_{x2}=80$, $L_{y1}=14$, and $L_{y2}=32$ in the units of $l_{\rm min}$. The time evolution of the $x$ component of the total force $F^{\rm tot}_{x}$ is shown, with the scale given on the right axis in orange. (C) Different protein clusters are shown in different colours on a $x-y$ plane from the top view during the polarity flip. We indicate the actin flow contribution from each cluster using a correspondingly coloured arrow. The total actin flow is indicated using a black arrow at the centre of mass of the vesicle. (D) A schematic diagram of the vesicle indicates that the CMC are coloured red for $x_i>x_{\rm CM}$, and blue for $x_i<x_{\rm CM}$. On the right, the dynamics of the mean curvature for the red and blue CMC is shown. The black dashed lines denote the reversals of the total force along the $x$ direction (B).}
    \label{fig:adhesion_confinement}
\end{figure}

\begin{figure}
    \centering
    \includegraphics[scale=0.5]{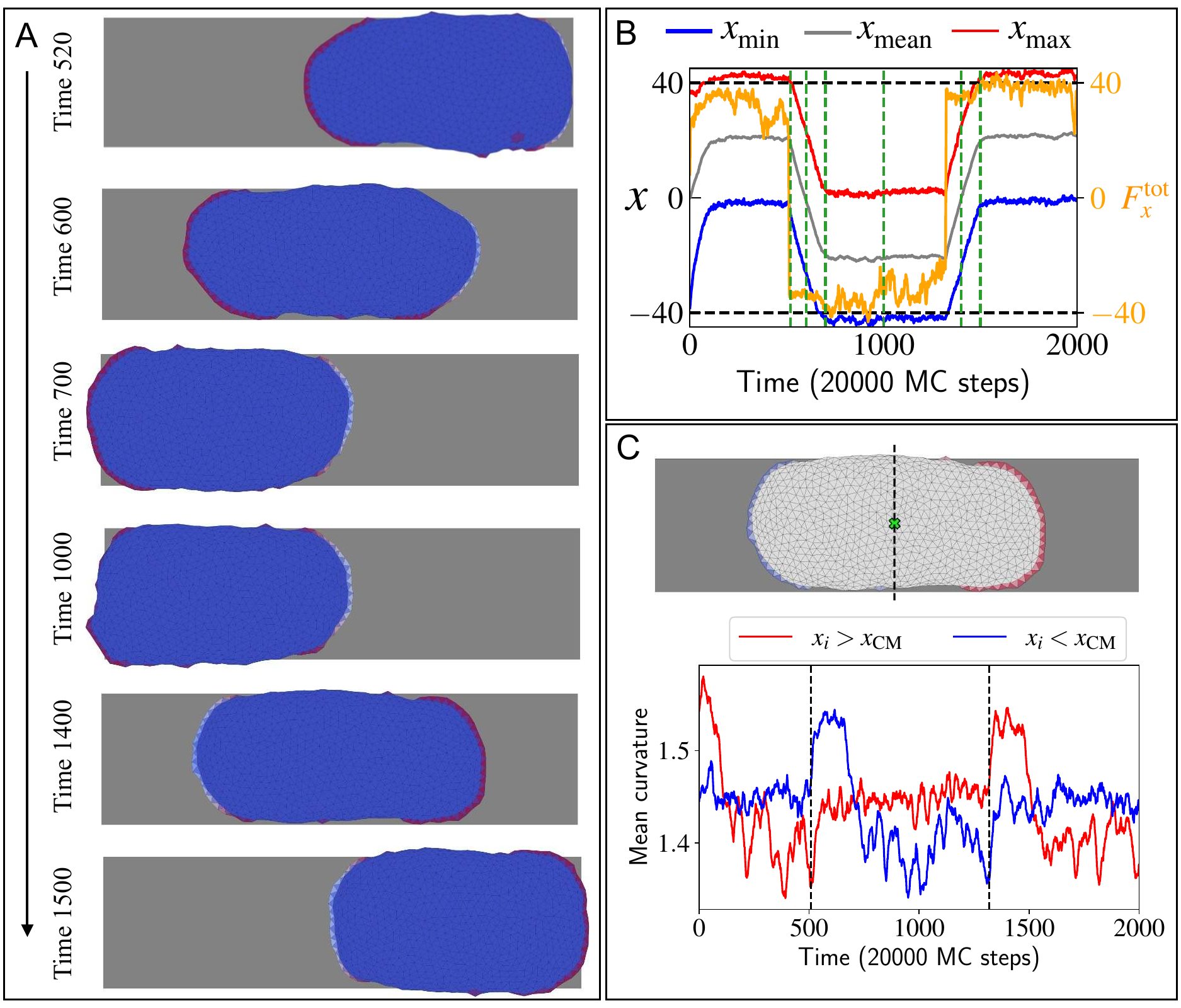}
    \caption{(A) The snapshots of the vesicle within the dumbbell-shaped confinement at time 520, 600, 700, 1000, 1400, and 1500 in the units of 20000 MC Steps during a complete oscillation on the rectangular adhesive substrate. The length and the width of the rectangle are 80$l_{\rm min}$ and 22$l_{\rm min}$ respectively. (B) The oscillation in $x$ coordinates of the vesicle over time. The $x_{\rm min}$, $x_{\rm mean}$, and $x_{\rm max}$ of all the vertices of the vesicle are shown in blue, grey, and red solid lines. The time evolution of the $x$ component of the total force $F^{\rm tot}_{x}$ is shown in the dual scale in orange. Green vertical dashed lines denote the time of the snapshots in (A). (C) On the top, a schematic diagram of the vesicle where the curved proteins are coloured with red if it's $x$ coordinate $x_i>x_{\rm CM}$, and with blue if $x_i<x_{\rm CM}$. The centre of mass is denoted by a lime-coloured marker. On the bottom, the mean curvature for the curved proteins for $x_i>x_{\rm CM}$ and $x_i<x_{\rm CM}$ in red and blue respectively. The black dashed lines denote the reversal of force along the $x$ direction.}
    \label{fig:rectangular_adhesion}
\end{figure}

\subsection{Dumbbell and rectangular adhesive patch}\label{SI_sec:adhesion_pattern}
We confined the vesicle movement by placing the "mini-cell" on an adhesive patch of a dumbbell and a rectangular shape. The vesicle feels the attractive adhesion only if the vertex of the vesicle is on the adhesive portion (denoted with a grey colour) of the substrate, otherwise, the energy of adhesion for the vertex is zero (white coloured). For the dumbbell-shaped adhesive patch (See Movie~\ref{mov:DBLL_adhesion}), we allowed the region mentioned in Eq.~\ref{eq:dumbbell_cond} for adhesion. The oscillation of the vesicle on the dumbbell-shaped adhesive pattern is shown in Fig.~\ref{fig:adhesion_confinement}. The oscillation of the vesicle on the dumbbell-shaped adhesive substrate is shown in Fig.~\ref{fig:adhesion_confinement}A. In Fig.~\ref{fig:adhesion_confinement}B, we showed the oscillation of $x$ positions of the vesicle and the total active force along $x$ axis. Fig.~\ref{fig:adhesion_confinement}C shows the breaking of the protein clusters and its contribution in actin flow and  finally, the repolarization in the opposite direction. The oscillation in curvature of the protein clusters belongs to right and left of the center of mass in red and blue respectively as shown in Fig.~\ref{fig:adhesion_confinement}D.

For the rectangular-shaped adhesive pattern, one can simply set the width $L_y=Ly_{1}=Ly_{2}$ and length $L_x=Lx_{2}$.  Fig.~\ref{fig:rectangular_adhesion} respectively. Our ``mini-cell'' model exhibits an oscillation (See Movie~\ref{mov:REC_adhesion}) along the length of the rectangular adhesive patch as shown in Fig.~\ref{fig:rectangular_adhesion}A. We showed the oscillation of the $x$ position (i.e., along the length of the rectangular adhesive patch) and the $x$ component of the total active force $F^{\rm tot}_x$ in Fig.~\ref{fig:rectangular_adhesion}B. In Fig.~\ref{fig:rectangular_adhesion}C, we showed the mean curvature of the protein sites of the vesicle for the cases $x_i>x_{\rm CM}$ and $x_i<x_{\rm CM}$ with red and blue respectively.\\

\subsection{Cell culture and imaging}
For the experimental recordings, we used non-axenic D. discoideum cells (DdB wildtype  background) that are deficient in NF1, a homologue of the human RasGAP Neurofibromin~\cite{Bloomfield2015}. They show a strongly increased tendency to switch to a highly polarized keratocyte-like mode of migration (so called fan-shaped motility)~\cite{Moldenhawer2022} that resembles the polar crescent-shaped model vesicles. The cell strain was transformed with an episomal plasmid encoding Lifeact-GFP and PHcrac-RFP (as described in~\cite{Flemming2020}. Cells were cultivated in 10 cm dishes with Sørensen's buffer (14.7 mM KH2PO4, 2mM Na2HPO4, pH 6.0) supplemented with 50 µM MgCl2, 50 µM CaCl2 (Sørensen's-MC buffer) and using G418 (5 µg/ml) and hygromycin (33 µg/ml) as selection markers. Klebsiella aerogenes with an OD600 of 20 were added to the solution in 1:10 volume to a final OD600 of 2. Before imaging, cells were washed with Sørensen's-MC buffer by centrifugation at 300 x g to remove any remaining bacteria. The resulting pellet was reconstituted in Sørensen's-MC buffer and cells were left to starve for 1 hour before infusing the cell solution in the microfluidic chip using a syringe pump. Imaging was performed without flow in the microfluidic chip, using a laser scanning microscope (LSM780, Zeiss, Jena) with a 488 nm Argon laser and a 561 nm diode laser. For the experiments shown in Figures 2 and 5, acquisition was done with a 20x objective at an interval of 20 s. For the experiments shown in Figure 6, acquisition was done with a 40x oil immersion objective, at an interval of 5 s. 

\subsection{Fabrication of microfluidic chips}
A silicon wafer coated with a 10 µm photoresist layer (SU-8 2010, Micro Resist Technology GmbH, Germany) was patterned by direct write lithography using a maskless aligner (µMLA, Heidelberg Instruments Mikrotechnik GmbH, Germany). Polydimethylsiloxane (PDMS, Sylgard 184, Dow Corning GmbH, Germany) at a ratio of 10:1 (base to curing agent) was poured into the microstructured wafer, degassed and cured for 2 h at 75°C. A PDMS block containing the microstructures was cut out and plasma bonded to a glass coverslip (\#1.5, 24 $\times$ 40 mm, Menzel Glaser). The microfluidic chip was rinsed extensively with Sørensen's-MC buffer before adding the cell solution.\\\\


\section{Movie }\label{mov:breaking_cluster}\textbf{Breaking of the unstable crescent vesicle, stabilized by UCSP}---{Crescent-shaped polar vesicle is very unstable. Simulation for four different $\beta=0,~0.1,~0.4,$ and $20$ in units of $D/k_BT$ in four columns respectively. We can see the stabilization of the vesicle as the coupling factor increases.\href{https://weizmann.box.com/s/4q8ixyh7xwv5llli25c3wnxhuj8730ph}{\textcolor{blue}{click here}}}

\section{Movie }\label{mov:breaking_low_kappa1}\textbf{Breaking of the unstable crescent vesicle, higher thermal fluctuation, $\beta=0~D/k_BT$}---{Crescent-shaped polar vesicle is very unstable. The bending rigidity of the membrane is set to $\kappa=15 k_BT$ which increases the thermal noise. It easily breaks the protein cluster and loses polarity.\href{https://weizmann.box.com/s/qnnyvmnvg9hnna4t9uun5ds6xmn57unj}{\textcolor{blue}{click here}}}

\section{Movie }\label{mov:breaking_low_kappa2}
\textbf{Breaking due to  higher thermal fluctuation, repolarize through UCSP $\beta=20~D/k_BT$}---{Crescent-shaped polar vesicle is very unstable. The bending rigidity of the membrane is set to $\kappa=15 k_BT$ which increases the thermal noise. It breaks the protein cluster, loses polarity partially and repolarizes due to the strong UCSP coupling $\beta=20~D/k_BT$.\href{https://weizmann.box.com/s/rqp7dp9zdrxjkzixo6w8e92t2ais1f9c}{\textcolor{blue}{click here}}}

\section{Movie }\label{mov:two_to_crescent_sim}\textbf{Transition from a two-arc shape to a crescent-shaped vesicle}---{A two-arc shaped vesicle $F=3k_BT l^{-1}_{\rm min},~E_{\rm ad}=1k_BT$, can make a spontaneous transition to a crescent-shaped polar vesicle with the UCSP coupling $\beta=10~D/k_BT$.\href{https://weizmann.box.com/s/hrcsvoa56fuz5e1eq2i2hblt1i2im4jx}{\textcolor{blue}{click here}}}

\section{Movie }\label{mov:transition_expt}\textbf{Experimental: Transition from two-arc to crescent and vice versa}---{
Timelapse recording of a \textit{D. discoideum} cell moving in between two PDMS barriers (rectangular shapes at the top and bottom of the field of view). The cell undergoes a transition from two-arc to crescent shape and vice versa. The time interval where the cell does not interact with the barrier and shows the transition, is presented in Fig. 2J, with t = 0 in the figure corresponding to t = 1000 s (timestamp 16:40) in the full timelapse. Green channel:  LifeAct-GFP, red channel: PHcrac-RFP.
\href{https://weizmann.box.com/s/w6vzj3fpwfyhao17kpsoi8t64eehz0f2}{\textcolor{blue}{click here}}}

\section{Movie }\label{mov:HW_beta0}
\textbf{Interaction with rigid barrier, No UCSP $\beta=0~D/k_BT$}---{A polar vesicle ($F=2k_BT l^{-1}_{\rm min},~E_{\rm ad}=3k_BT$) hitting a rigid barrier $\xi=90^o$ and breaks into a non-polar two-arc shape as UCSP coupling $\beta=0~D/k_BT$ is absent.\href{https://weizmann.box.com/s/sxd4ua2ff1xxewfgdqim5965xs7px0ot}{\textcolor{blue}{click here}}}

\section{Movie }\label{mov:HW_beta15}\textbf{Interaction with rigid barrier, UCSP $\beta=15~D/k_BT$}---{A polar vesicle ($F=2k_BT l^{-1}_{\rm min},~E_{\rm ad}=3k_BT$) hitting a rigid barrier $\xi=90^o$ and repolarize to a polar crescent shaped vesicle as UCSP coupling $\beta=15$ is strong enough.\href{https://weizmann.box.com/s/onepu4lmcgcg3rk4fy4ic7h3vmpk7tzk}{\textcolor{blue}{click here}}}

\section{Movie }\label{mov:HW_scatter_sim}\textbf{Scattering from the rigid barrier in large angle}---{Scattering of the vesicle on the rigid barrier. We used parameters $F=2k_BT l^{-1}_{\rm min},~E_{\rm ad}=3k_BT$,~$\beta=6~D/k_BT$.\href{https://weizmann.box.com/s/w3xobwjyr5hp1hfbhn0wr9lvv6nplit7}{\textcolor{blue}{click here}}}

\section{Movie }\label{mov:HW_slide_with_wall_expt}\textbf{Experimental: Sliding along the wall}---{
Timelapse recording of a \textit{D. discoideum} cell sliding along the wall of a PDMS barrier. The corresponding trajectory and snapshots are shown in Fig. 5C. Green channel:  LifeAct-GFP, red channel: PHcrac-RFP.\href{https://weizmann.box.com/s/89r5fxt2d6p90xqlkfnduejd5syvgbtk}{\textcolor{blue}{click here}}}

\section{Movie }\label{mov:HW_scatter_expt}\textbf{Experimental: Scattering in large angle}---{
Timelapse recording of \textit{D. discoideum} cells moving in between two PDMS barriers (rectangular shapes at the top and bottom of the field of view). Timestamp is shown in mm:ss. Green channel:  LifeAct-GFP, red channel: PHcrac-RFP.\href{https://weizmann.box.com/s/rm3pau56yrqs9zza2lry29nlz2edfawx}{\textcolor{blue}{click here}}}

\section{Movie }\label{mov:SW_beta0}\textbf{Interaction with soft barrier, No UCSP $\beta=0~D/k_BT$}---{
A polar vesicle ($F=2k_BT l^{-1}_{\rm min},~E_{\rm ad}=3k_BT$) hitting a soft barrier $\xi=90^o$. The vesicle penetrates the barrier but finally breaks into a non-polar two-arc shape as UCSP coupling $\beta=0~D/k_BT$ is absent.\href{https://weizmann.box.com/s/0a7h51ufvv7f4zca5njljz9c8e6v3xsu}{\textcolor{blue}{click here}}}

\section{Movie }\label{mov:SW_beta15}\textbf{Interaction with soft barrier, UCSP $\beta=15~D/k_BT$}---{A polar vesicle ($F=2k_BT l^{-1}_{\rm min},~E_{\rm ad}=3k_BT$,~$\beta=15~D/k_BT$) hitting a soft barrier $\xi=90^o$, is moving along the wall with a partial penetration for a longer time.\href{https://weizmann.box.com/s/a302pnbnh8kl984s8p19yz68qm4cvpfi}{\textcolor{blue}{click here}}}

\section{Movie }\label{mov:TT_beta0}\textbf{Interaction with triangular tip, No UCSP $\beta=0~D/k_BT$ }---{A polar vesicle is placed in front of a triangular tip. The leading-edge protein cluster breaks and remains non-polar as the UCSP coupling $\beta=0~D/k_BT$ is absent.\href{https://weizmann.box.com/s/3fpm7krg3xgurw12m5b9deusj4xwzsow}{\textcolor{blue}{click here}}}

\section{Movie }\label{mov:TT_beta20}\textbf{Interaction with triangular tip, UCSP $\beta=20~D/k_BT$}---{A polar vesicle is placed in front of a triangular tip. The leading-edge protein cluster breaks and loses its polarity and repolarizes itself as the UCSP coupling $\beta=20~D/k_BT$ is strong.\href{https://weizmann.box.com/s/44c89nhg93c6ijsktfvadjz6jx92ldyb}{\textcolor{blue}{click here}}}

\section{Movie }\label{mov:TT_expt}\textbf{Experimental: Interaction with triangular tip}---{Timelapse recording of a \textit{D. discoideum} cell interacting with a triangular PDMS barrier. The corresponding snapshots are shown in Fig. 6D, bottom. Timestamp is shown in mm:ss. Green channel:  LifeAct-GFP, red channel: PHcrac-RFP.\href{https://weizmann.box.com/s/0ss3vb54ysf4dn5colsmcdi98cxmj2cd}{\textcolor{blue}{click here}}}

\section{Movie }\label{mov:DBLL_rigid}\textbf{Oscillation on a Dumbbell patterned rigid barrier}---{Oscillation of the vesicle within the dumbbell-shaped rigid barrier. We used parameters $F=3k_BT l^{-1}_{\rm min},~E_{\rm ad}=3k_BT$,~$\beta=20~D/k_BT,~L_{x1}=36l_{\rm min},~L_{x2}=78l_{\rm min},~L_{y1}=16l_{\rm min},~L_{y2}=32l_{\rm min}$.\href{https://weizmann.box.com/s/ry4bchqfi51fmfr7gz9xcl1bn43h6907}{\textcolor{blue}{click here}}}

\section{Movie }\label{mov:DBLL_adhesion}\textbf{Oscillation on a Dumbbell patterned adhesion}---{Oscillation of the vesicle on the dumbbell-shaped adhesive pattern. We used parameters $F=3k_BT l^{-1}_{\rm min},~E_{\rm ad}=3k_BT$,~$\beta=20~D/k_BT,~L_{x1}=40l_{\rm min},~L_{x2}=80l_{\rm min},~L_{y1}=14l_{\rm min},~L_{y2}=32l_{\rm min}$.\href{https://weizmann.box.com/s/g4bsmkkof21mt0hhqhvqlxov109hmwxu}{\textcolor{blue}{click here}}}

\section{Movie }\label{mov:REC_adhesion}\textbf{Oscillation on a rectangular adhesive pattern}---{Oscillation of the vesicle on the rectangular-shaped adhesive pattern. We used parameters $F=3k_BT l^{-1}_{\rm min},~E_{\rm ad}=3k_BT$,~$\beta=20~D/k_BT,~L_x=80l_{\rm min},~L_y=22l_{\rm min}$.\href{https://weizmann.box.com/s/nt8czxu02j4xxnffgakvfcdl83mwdjmt}{\textcolor{blue}{click here}}}

\bibliographystyle{abbrv}
\bibliography{shubhadeep_biophys}